\documentclass[aps,12pt,tightenlinesletterpaper,superscriptaddress,secnumarabic,nofootinbib]{revtex4}
\usepackage[dvips]{graphicx}

\DeclareRobustCommand{\baselinestretch{1.0}}

\usepackage{amsmath}
\usepackage{bm}
\usepackage{epsfig}
\usepackage{verbatim}
\usepackage{psfrag}
\usepackage{bm}
\usepackage{cancel}
\usepackage[dvips]{hyperref}

\allowdisplaybreaks[1]

\newcommand{\lsim}{\buildrel < \over {_\sim}}
\newcommand{\gsim}{\buildrel > \over {_\sim}}
\newcommand{\be}{\begin{equation}}
\newcommand{\ee}{\end{equation}}
\newcommand{\bea}{\begin{eqnarray}}
\newcommand{\eea}{\end{eqnarray}}
\newcommand{\ba}{\begin{array}}
\newcommand{\ea}{\end{array}}

\newcommand{\lp}{\left(}
\newcommand{\rp}{\right)}
\newcommand{\lb}{\left[}
\newcommand{\rb}{\right]}

\newcommand{\of}[1]{\left(#1\right)}
\newcommand{\eqnref}[1]{$\left(\ref{#1}\right)$}
\newcommand{\msbar}{\overline{\mathrm{MS}}}
\newcommand{\unit}[1]{\,\text{#1}}
\newcommand{\comma}{,\,}
\newcommand{\loopfac}{\frac{1}{16\pi^2}}
\newcommand{\veff}{V_\mathrm{eff}}

\newcommand{\singlettree}{V(h,S)^{\lp \mathrm{tree}\rp}}
\newcommand{\singletoneloop}{V(h,S)^{\lp\mathrm{ 1\,loop}\rp}}

\newcommand{\svev}{\langle S\rangle}

\newcommand{\ofmu}{\lp\mu\rp}
\newcommand{\ofv}{\lp v_0\rp}
\newcommand{\model}{$Z_2$xSM}

\begin{document}
\hfill NPAC-09-13
\title{\Large Vacuum Stability, Perturbativity, and\\ Scalar Singlet Dark Matter}
\vspace{4.0cm}
\author{Matthew~Gonderinger}
\email{gonderinger@wisc.edu}
\author{Yingchuan~Li}
\email{yli@physics.wisc.edu}
\author{Hiren~Patel}
\email{hhpatel@wisc.edu}
\affiliation{
{\small Department of Physics, University of Wisconsin-Madison} \\
{\small Madison, WI 53706, USA}}
\author{Michael~J.~Ramsey-Musolf}
\email{mjrm@physics.wisc.edu}
\affiliation{
{\small Department of Physics, University of Wisconsin-Madison} \\
{\small Madison, WI 53706, USA}}
\affiliation{
{\small Kellogg Radiation Laboratory, California Institute of Technology} \\
{\small Pasadena, CA, 91125, USA}}

\begin{abstract}
We analyze the one-loop vacuum stability and perturbativity bounds on a singlet extension of the Standard Model (SM) scalar sector containing a scalar dark matter candidate. We show that the presence of the singlet-doublet quartic interaction relaxes the vacuum stability lower bound on the SM Higgs mass as a function of the cutoff and lowers the corresponding upper bound based on perturbativity considerations. We also find that vacuum stability requirements may place a lower bound on the singlet dark matter mass for given singlet quartic self coupling, leading to restrictions on the parameter space consistent with the observed relic density. We argue that discovery of a light singlet scalar dark matter particle could provide indirect information on the singlet quartic self-coupling. 
\end{abstract}
\maketitle
\section{Introduction}
\label{sec:intro}

The search for the mechanism of electroweak symmetry-breaking (EWSB) is one of the primary aims of on-going experiments at the Tevatron and future experiments at the Large Hadron Collider. Until a Standard Model (SM) Higgs boson is discovered at either collider, it is by no means assured that EWSB occurs through a low-energy scalar sector that contains only a single, complex SU(2$)_L$ doublet. The simplest variant of the latter scenario involves the presence of an additional scalar degree of freedom (real or complex) that carries no SM gauge charges. If it exists, a scalar singlet could be the low-energy remnant of a more complex higher energy model. Depending on the nature of the scalar potential, a real or complex singlet could provide a viable candidate for the observed relic abundance of cold dark matter \cite{Barger:2008jx,Bento:2001yk,Silveira:1985rk,McDonald:1993ex,Burgess:2000yq,
He:2008qm,McDonald:2001vt,Bento:2000ah,Barger:2007im,O'Connell:2006wi} and/or give rise to a strong first-order phase transition during the era of EWSB as needed for electroweak baryogenesis \cite{Profumo:2007wc,Espinosa:1993bs,Anderson:1991zb}.  Its existence could be verified experimentally through the discovery of one or more mixed SU(2$)_L$ doublet-singlet states in colliders \cite{O'Connell:2006wi,Barger:2007im}, the observation of a distinctive azimuthal correlation between outgoing jets in weak boson fusion \cite{Eboli:2000ze,Barger:2006sk,Davoudiasl:2004aj}, and a signal in the next generation of dark matter direct detection experiments \cite{Barger:2007im,Brink:2005ej,Carroll:2009dw}. 

If a scalar singlet is discovered, a combination of the foregoing experiments could also provide information about the parameters of the scalar potential, particularly those that involve both the doublet and singlet degrees of freedom. Probing the singlet self-interactions will be more difficult. This problem is particularly interesting if the scalar singlet provides a dark matter candidate, as the nature of the dark sector and its interactions is presently unknown. Bounds on the scalar singlet quartic self-coupling can be derived from the dynamics of the Bullet cluster \cite{McDonald:2007ka} and from observations of galactic dark matter halos \cite{Bento:2000ah,McDonald:2001vt,Spergel:1999mh,McDonald:2007ka,Bento:2001yk} if the scalar is very light ($M_S$ on the order of 1 to 100 MeV) but for heavier masses there presently exist no limits. 

It is also interesting to ask, if the SM scalar sector is accompanied by a singlet, at what scale $\Lambda$ should one expect the high energy desert to be populated with additional degrees of freedom? At one-loop order, the absence of a Landau pole in the scalar self-interactions should provide a rough guide as to this scale. Since the value of $M_h$ fixes the value of the Higgs quartic coupling at an input scale for its renormalization group (RG) evolution, 
the discovery of a heavy Higgs could provide one indication of a maximal value for $\Lambda$. On the other hand, if a relatively light scalar is discovered, $\Lambda$ cannot in general be too large if the vacuum associated with EWSB is to be stable. The corresponding considerations have been analyzed extensively in the case of the SM (more recently in Refs.~\cite{Hambye:1996wb,Casas:1996aq}), with an allowed range of $55\ \mathrm{GeV} \lsim M_h \lsim (500-800)\ \mathrm{GeV}$ for $\Lambda\sim 1$ TeV and a correspondingly narrow range of $150\ \mathrm{GeV} \lsim M_h \lsim 200\ \mathrm{GeV}$ for $\Lambda$ at the GUT scale (neglecting theoretical and parametric uncertainties)\footnote{As discussed below, the two-loop analysis of Ref.~\cite{Hambye:1996wb} ties the upper bound to the fixed point of the two-loop $\beta$-function; for large $\Lambda$, the results are not too different from those obtained using the one-loop Landau pole.}. 

To our knowledge, the corresponding expectations for singlet extensions of the SM have not been thoroughly studied, although partial analyses were performed in Refs.~\cite{Clark:2009dc,Lerner:2009xg}.  These works considered vacuum stability and perturbativity constraints on a singlet extension of the SM in the context of inflation models which identified the inflaton with either the Higgs or the scalar singlet.  The authors included non-minimal gravitational coupling between the Ricci scalar and the inflaton, {\em i.e.} the Higgs or singlet, and showed that cosmological observables place restrictions on the parameter space which can be stronger than the vacuum stability and perturbativity limits.  In this work, we do not assume that either the Higgs or the singlet serves as the inflaton and so we do not encounter constraints imposed by inflation.

In what follows, we address these issues for the scalar singlet extension of the SM by analyzing the one-loop vacuum stability and perturbativity constraints on the theory. For simplicity, we restrict our attention to the case of a real scalar singlet $S$ for a potential possessing a $Z_2$ symmetry -- henceforth denoted the \model~-- and require that the singlet vacuum expectation value (vev) $\langle S \rangle=0$ in order to provide for a stable singlet that may be a dark matter candidate. We first find that, in contrast to the SM case, a perturbativity criterion based on the avoidance of a Landau pole at large $\Lambda$ can lead to an unreliable analysis of the vacuum structure.  Consequently, a more stringent one-loop perturbativity requirement must be imposed.  We then show that for a given value of $\Lambda$,  the presence of the quartic $S$-$H$ coupling, henceforth denoted $a_2$, lowers both the upper and lower bounds on $M_h$. For given values of $M_h$ and $M_S$, this coupling must generally be sufficiently large in order that the relic density of singlet scalars does not over-saturate the observed cold dark matter relic density. In this scenario, then, the discovery of a relatively heavy Higgs would imply a lower scale for the onset of new physics than in the SM. These considerations are particularly pronounced for larger $\Lambda$. 

We also find that requiring that $\langle S \rangle=0$ be the true vacuum can have interesting implications for the singlet scalar self-coupling, denoted $b_4$ below, particularly when consistency with the observed relic density is imposed. For given values of $M_h$, $M_S$, and $\Lambda$, these considerations can imply a lower bound on the magnitude of this self coupling. Alternately, fixing the magnitude of the latter can lead to a lower bound on the singlet dark matter mass. To illustrate,  for $\Lambda=1$ TeV and $M_h=120$ GeV, a vanishing self-coupling implies $M_S \gsim 50$ GeV. Increasing $b_4$ reduces this bound while increasing $\Lambda$ raises it. For $\Lambda > 10^{9}$ GeV and $M_h=120$ GeV, there exists no region of the $\of{M_S\comma b_4\comma a_2}$ parameter space that can accommodate the relic density. In short, assumptions about the magnitude of the scalar self coupling and the scale of new physics can impact the degree to which a scalar singlet can explain the DM relic density. 

In the remainder of the paper, we present the analysis leading to these conclusions. In Section \ref{sec:potential}, we review the model potential. Section \ref{sec:vacstab} gives our analysis of the one-loop, RGE-improved effective potential and the criteria we choose to maintain a perturbative theory.  We show illustrative applications of our analysis to the model parameter space and discuss the implications of these results in Section \ref{sec:disc}.  We summarize our work in Section \ref{sec:concl}.

\section{Scalar Potential}
\label{sec:potential}

The most general renormalizable potential involving the SM Higgs doublet $H$ and the singlet $S$ is \cite{O'Connell:2006wi,Profumo:2007wc,Barger:2007im}
\begin{equation}
\label{Scalar_singlet_tree_level_potential}
	\singlettree = m^2H^{\dagger}H + \frac{\lambda}{6}\lp H^{\dagger}H\rp^2 + a_1SH^\dagger H
		+ a_2S^2H^\dagger H + \frac{b_2}{2}S^2 + \frac{b_3}{3}S^3 + \frac{b_4}{4}S^4+\Omega\ \ \ ,
\end{equation}
where $m^2<0$ and where we have eliminated a possible linear term in $S$ by a constant shift, absorbing the resulting $S$-independent term in the vacuum energy $\Omega$. To set our notation\footnote{For a comparison of our notation and normalizations with those used in the literature, see the Appendix.} we take
\begin{equation}
\label{Standard_model_higgs_doublet}
	H = \frac{1}{\sqrt{2}}\lb\begin{array}{c} \phi_1+i\phi_2 \\ h+i\phi_4\end{array}\rb \ \ \ .
\end{equation}
We require that the minimum of the potential occur at $h=246\,\text{GeV}\equiv v_0$. Fluctuations around this vacuum expectation value (vev) are the SM Higgs boson and the $\phi_i$ are the would-be Goldstone bosons that become the longitudinal components of the electroweak gauge bosons. 

For the case of interest here for which $S$ is stable and may be a dark matter candidate, we impose a $Z_2$ symmetry on the model, thereby eliminating the $a_1$ and $b_3$ terms. We also require that the true vacuum of the theory satisfy $\svev=0$, thereby precluding mixing of $S$ and the SM Higgs boson and the existence of cosmologically problematic domain walls. In this case, the tree-level minimization conditions give
\be
\label{Tree_level_vev}
	v_0 = \sqrt{-\frac{6m^2}{\lambda}}\ \ \ ,
\ee
with masses for the doublet and singlet scalars
\bea
\label{eq:mhtree}
m_{h}^2& = & \frac{\lambda v_0^2}{3} \\
\label{eq:mstree}
m_S^2 & = & b_2 + a_2 v^2_0 \ \ \ ,
\eea
where we have utilized the tree-level relation (\ref{Tree_level_vev}) to eliminate $m^2$ in favor of $v_0^2$.

For purposes of analyzing vacuum stability, inclusion of radiative corrections is essential. We begin with
the one-loop effective potential, $\veff(h,S)$, whose computation requires knowledge of the field-dependent masses for the scalar, fermionic, and gauge boson degrees of freedom, denoted $m_j(h,S)$. The one-loop contribution, renormalized in $\msbar$, is given by
\begin{equation}
\label{One_loop_potential_formula}
	\singletoneloop = \sum_j\frac{n_j}{64\pi^2}m_j^4\lp h,S\rp\lb\log{\frac{m_j^2\lp h,S\rp}{\mu^2}} - c_j\rb\ \ \ ,
\end{equation}
where $j$ runs over all degrees of freedom in the model and where $h$ and $S$ now denote the classical fields. For the fermions, we retain only the top quark contribution since the contributions from the remaining fermions are suppressed by their Yukawa couplings and have a negligible impact on our analysis. For the neutral scalars, allowing for both classical fields $h$ and $S$ to be non-vanishing away from the minimum of the potential leads to mixing between these degrees of freedom associated with the off-diagonal element in the mass-squared matrix
\be
\label{eq:msquared}
{\bm M}^2(h,S)= \left(\begin{array}{cc}
m^2+\frac{\lambda}{2}h^2+a_2 S^2 & 2a_2 h S \\
2a_2 h S & b_2+3b_4 S^2 +a_2 h^2 
\end{array}\right)\equiv\lp\begin{array}{cc}
m_h^2&m_{hS}^2\\m_{hS}^2&m_S^2
\end{array}\rp\ \ \ .
\ee
The corresponding field-dependent eigenvalues are
\be\label{eq:plus_minus_mass_evals}
m_{\pm}^2=\frac{1}{2}\lb\lp m_h^2+m_S^2\rp\pm\sqrt{\lp m_h^2-m_S^2\rp^2+4m_{hS}^4}\ \rb\ \ \ .
\ee
The remaining field-dependent masses, constant factors $c_j$, and degeneracy factors $n_j$ are listed in the Appendix. 

As seen in Eq.~\eqnref{One_loop_potential_formula}, the one-loop contribution to the potential depends explicitly on the arbitrary unphysical renormalization parameter (or 't Hooft scale) $\mu$. In order to ensure that the full potential is scale independent, we introduce the running parameters $\lambda\rightarrow\lambda\lp\mu\rp,\,m^2\rightarrow m^2\lp\mu\rp,\,h\rightarrow h\lp\mu\rp,$ {\em etc.} that also depend on the scale $\mu$.  The full potential through one-loop is given by 
\be
\label{eq:vefffull}
\veff(h,S)=\singlettree+\singletoneloop
\ee
with
\be
\label{eq:treerun}
\singlettree=\frac{m^2(\mu)}{2}h^2+\frac{\lambda(\mu)}{24} h^4 +\frac{a_2(\mu)}{2} S^2 h^2 +\frac{b_2(\mu)}{2}S^2+
\frac{b_4(\mu)}{4} S^4 +\Omega(\mu)\ \ \ ,
\ee
where we have suppressed the $\mu$-dependence of the fields $h=h(\mu)$ and $S=S(\mu)$ for simplicity .

The running parameters are obtained from the beta and gamma functions and the corresponding renormalization group equations (RGEs): 
\begin{align}
	&\beta_X \equiv \mu\frac{dX}{d\mu}\\
	&\gamma_Y \equiv -\frac{\mu}{Y}\frac{dY}{d\mu} = \frac{\mu}{2}\frac{d\log{Z_Y}}{d\mu}
\end{align}
for a given coupling or mass parameter X and field Y ($Z_Y$ is the wave function renormalization for Y).  The beta functions for the gauge couplings, $g$, $g'$, and $g_3$, and the top quark Yukawa coupling, $y_t$, are unaffected to one-loop order by the addition of the singlet; thus they are the same as in the Standard Model and can be found in Ref.~\cite{Ford:1992mv}.  Additionally, the wave function renormalizations for the Higgs and the singlet do not receive any one-loop scalar contributions.  As a result, $\gamma_h$ remains the same as in the Standard Model (also in Ref.~\cite{Ford:1992mv}), and $\gamma_S=0$.  

All that remains is to find the beta functions for the scalar potential parameters.  These are determined by requiring scale independence of the effective potential in Eq.~\eqnref{eq:vefffull} to one-loop order:
\begin{equation}\label{One_loop_scale_independent_potential}
	\mu\frac{d\veff(h,S)}{d\mu} =    
		\mu\frac{d\singlettree}{d\mu} + \mu\frac{\partial\singletoneloop}{\partial\mu} = 0
\end{equation}
Using Eqs.~\eqnref{eq:treerun} and \eqnref{One_loop_potential_formula}, the definitions for $\beta_X$ and $\gamma_Y$ above, and the known results for $\gamma_h$ and $\gamma_S$, we use this scale cancellation to obtain the beta functions for $\lambda$, $a_2$, $b_4$, $m^2$, and $b_2$ by comparing terms of the same degree in $h$ and $S$.  The running of the vacuum energy is found from terms that are independent of the fields. We have independently verified our results, given in the Appendix, by explicitly computing the one-loop $\msbar$ renormalization of the quadratic and quartic terms in the potential and using the scale-independence of the bare parameters\footnote{Our results for the beta functions also match those in Refs.~\cite{Clark:2009dc,Lerner:2009xg}.}.

As boundary conditions for the running gauge couplings, we take
\begin{align}
	&g\of{M_Z} = 0.65\nonumber\\
	&g'\of{M_Z} = 0.358\nonumber\\
	&g_3\of{M_Z} = 1.21\ \ \ .\nonumber
\end{align}
For the top Yukawa coupling boundary condition, we follow the choice of Ref.~\cite{Casas:1996aq}:
\begin{equation*}
	y_t\of{M_t} = \sqrt{2}\ \frac{m_t\of{M_t}}{\xi\of{M_t}v_0}\nonumber \ \ \ ,
\end{equation*}
where the running top mass, $m_t\of{\mu}$, is related to the top quark pole mass, $M_t$, by
\begin{equation*}
	M_t=\lp 1+\frac{4}{3}\frac{\alpha_s\of{M_t}}{\pi}\rp m_t\of{M_t}\ \ \ ,
\end{equation*}
and the factor $\xi\of{M_t}$ is explained below.

For the quartic couplings, we choose the input scale $\mu_0=M_Z$ and will vary $\lambda(M_Z)$, $b_4(M_Z)$, and $a_2(M_Z)$ according to various criteria discussed in Section \ref{sec:vacstab} below.   The quadratic mass-squared parameter $m^2(\mu)$ will be related to the Higgs boson pole mass using the minimization conditions on $\veff$. At one-loop order, it is convenient to relate these quantities at $\mu=v_0$, so we will take $v_0$ as the input scale for both running quadratic parameters\footnote{Note that the beta functions for the couplings are independent of the $b_2$ and $m^2$ parameters, allowing them to be solved independently without the minimization conditions.  Once we have determined the running of the couplings, and because $\beta_{b_2}$ depends linearly on $m^2(\mu)$, we can choose any scale for our boundary conditions -- $v_0$ is the most convenient choice because of the minimization conditions.}.

We take as the solution for the Higgs field RGE
\begin{equation}\label{Running_higgs_field}
	h\lp\mu\rp = \xi\lp\mu\rp h = \exp{\lp-\int_{M_Z}^\mu \gamma_h\lp\mu '\rp d\mu '/\mu '\rp} h
\end{equation}
with boundary condition
\begin{equation*}
	\xi\of{M_Z} = 1
\end{equation*}
so that
\begin{equation*}
	h\of{M_Z} \equiv h\ \ \ .
\end{equation*}
The running of the classical Higgs field $h\ofmu$ in \eqnref{Running_higgs_field} is very weak -- $\xi\ofmu$ deviates from one by less than 10\% over the range $M_Z\leq\mu\leq 10^{19}~\mathrm{GeV}$ (assuming there are no Landau poles present in this range).  Meanwhile, $S$ receives no scale-dependence at one-loop order since it does not couple to gauge fields and since quartic scalar interactions generate no wavefunction renormalization.  

We note that our choice of input scale for the quartic and quadratic parameters would not be optimal for a two-loop analysis, since the corresponding one-loop matching corrections can be unacceptably large. As discussed in Ref.~\cite{Hambye:1996wb}, a suitable choice is $\mu_0=\mathrm{max}\{M_t, M_h\}$. Since we do not encounter any loop level matching conditions in our one-loop $\veff$ analysis, the only impact of varying our input scale between $M_Z$ and $M_t$ is via the running parameters and the explicit one-loop term in $\veff$, and the corresponding numerical impact is not significant. Consequently, we choose an input scale for each parameter that renders the explicit logarithms in Eq.~\eqnref{One_loop_potential_formula} small and that is most convenient for fixing the corresponding boundary condition for the running.

Finally, we comment on a minor difference between our treatment of the scale-dependence and that of Ref.~\cite{Casas:1996aq}. The authors of that work identified an optimum scale $\mu^\ast$ that minimized the scale dependence of the effective potential and performed their minimization of the potential at that scale $\mu^\ast$. In the present work, no such choice of a special scale is made; since we include both terms in \eqnref{eq:vefffull}, the effective potential is scale-independent to one-loop order with the residual scale dependence being higher order. When analyzing the behavior of $\veff$ at large values of the field as needed for vacuum stability considerations, we will choose $\mu$ so as to minimize the explicit one-loop logarithms, resumming them implicitly through the running parameters in the first term of Eq.~\eqnref{eq:vefffull}.

\section{Vacuum Stability \& Perturbativity}
\label{sec:vacstab}

It is well-known that in the SM, $\veff(h)$ can develop a minimum at large values of $h$ ({\em i.e.,} $h>>v_0$) that is deeper than the electroweak minimum\footnote{For a comprehensive review, see Ref.~\cite{Sher:1988mj} and the references therein; also, see Refs.~\cite{Lindner:1985uk,Lindner:1988ww,Sher:1993mf,Casas:1994us,Casas:1994qy,Casas:1996aq}.}. This situation arises because of the large, negative top Yukawa contribution to $\beta_\lambda$ that causes $\lambda$ to run negative at large scales. In order to avoid the presence of this second, deeper minimum at scales below the cutoff $\Lambda$ of the effective theory, one requires a sufficiently large value of $\lambda$ at the input scale (here taken to be $\mu_0=M_Z$) so that the turnover of $\veff$ occurs at sufficiently large scales. On the other hand, a value of $\lambda(\mu_0)$ that is too large will lead to the presence of a Landau pole below $\Lambda$. These considerations lead, respectively, to the vacuum stability and triviality bounds on $M_h$, whose value is governed by  $\lambda$ at the electroweak scale. 

It is, of course, possible that the SM vacuum is metastable with respect to the second minimum at large $h$ with a lifetime longer than the age of the Universe. If during the process of electroweak symmetry-breaking the Universe first lands in the finite temperature electroweak vacuum, it may remain there as the temperature decreases even if the large $h$ minimum eventually becomes deeper. Recent works such as Refs.~\cite{Ellis:2009tp,Isidori:2001bm} have shown that allowing for a sufficiently long-lived electroweak vacuum can weaken the lower bounds on the Higgs mass by as much as $\sim 25\unit{GeV}$ (this change was roughly uniform for cutoff scales $10^{10}\unit{GeV}\leq\Lambda\leq 10^{19}\unit{GeV}$).  A careful analysis of the metastability bounds requires computing the one-loop corrections to the vacuum tunneling probability, which is beyond the scope of the present work; we restrict our attention to the more conservative criterion of absolute vacuum stability at $T=0$.

To obtain the  one-loop stability bound in the \model, we first analyze the behavior of the potential along the $h$-direction ($S=0$), and determine the value of $\lambda(M_Z)$ such that the potential does not develop an absolute minimum for $\Lambda> h >> v_0$. We choose the values of the singlet parameters at the input scale such that no minima occur away from the $h$ axis, and subsequently study the impact of departing from this choice. In order to obtain the most reliable one-loop analysis, we minimize the impact of large logarithms in the explicit one-loop contribution to $\veff$  by choosing $\mu=h(M_Z)$ (recall that $h(\mu)$ varies gently with $\mu$). Since the field-dependent masses along the $h$-direction are also proportional to $h$, this procedure leads to the appearance of explicit one-loop logarithms having order one arguments, while resumming the large logarithms through the running parameters  in $V\of{h,S}$. We then study $\veff$ as a function of $h(M_Z)\equiv h$. An alternate choice, such as choosing $\mu=M_Z$, would lead to explicit, un-resummed large logarithms for large values of $h$, rendering our one-loop analysis less reliable. 

To obtain the electroweak/dark matter minimum, we choose $m^2(\mu)$ such that the minimum of the potential occurs for $h\ofv\simeq h\of{M_Z}=v_0$,
\begin{equation}
\label{One_loop_minimization_higgs}
	\left.\frac{\partial\veff}{\partial h}\right|_{h=v_0,\,S=0,\,\mu=v_0} = 0\ \ \ ,
\end{equation}
with positive eigenvalues for the matrix of second derivatives (see below).
Except for $m^2\ofv$, all quantities in Eq.~\eqnref{One_loop_minimization_higgs} are known numerical values (once the RGEs for the couplings and the Higgs field have been solved). We subsequently plot $\veff$ as a function of $h(M_Z)$, with $\mu=h(M_Z)$ and determine whether or not a second, deeper minimum occurs below $\Lambda$. For given values of $a_2(M_Z)$, $b_4(M_Z)$, and $b_2(v_0)$, we repeat the procedure for successively smaller values of $\lambda(M_Z)$ until the appearance of a second, deeper minimum -- thereby obtaining the minimum value of $\lambda(M_Z)$ consistent with absolute stability up to the scale $\Lambda$. 

To arrive at a corresponding upper bound on $\lambda(M_Z)$, we begin by fixing $b_4(M_Z)$, $a_2(M_Z)$, and $b_2(v_0)$ and solving for the value of $\lambda(M_Z)$ that leads to a Landau pole at the cutoff $\Lambda$. We observe that the use of this triviality criterion to establish an upper bound on $\lambda(M_Z)$ is somewhat subjective. An alternate possibility is to require that the theory remain perturbative up to the cutoff since the effective potential has been computed using perturbation theory.  At one-loop order, one minimally must avoid the occurrence of the Landau pole below $\Lambda$; doing so, however, does not guarantee that the theory remains perturbative below the cutoff.  In fact, we find that using this triviality criterion leads to the presence of unstable minima at large scales ({\em i.e.,} large radii in $h$-$S$ field space) because the couplings in the loop corrections become non-perturbative.  Whether or not the existence of such minima is a physical effect is therefore suspect.  To avoid such problems, we choose to adopt more stringent perturbative restrictions on the running coupling parameters at the scale $\Lambda$ rather than rely on the triviality criterion to arrive at an upper bound.

Detailed studies of the perturbativity requirements in the SM have been carried out using next order (two loop) $\beta$-function analyses, resummation techniques, and lattice methods. At two-loop order,  the Higgs quartic coupling reaches a fixed point, $\lambda\equiv\lambda_\mathrm{FP}$ (corresponding to $\beta_\lambda =0$), rather than a Landau pole.  According to the analysis of Ref.~\cite{Hambye:1996wb}, the input value $\lambda(M_Z)$ leading to $\lambda(\Lambda)=\lambda_\mathrm{FP}$ is significantly smaller than the value that gives a Landau pole at $\Lambda$ for small values of the cutoff, with the difference decreasing as the cutoff is increased. In addition, the work of Ref.~\cite{Riesselmann:1996is} suggests that for $\lambda(\Lambda)< \lambda_\mathrm{FP}/4$ the theory remains perturbative up to $\Lambda$, while for $\lambda(\Lambda) < \lambda_\mathrm{FP}/2$, one may be reaching the limit of perturbative behavior. Thus, we expect that for the lower range of our cutoff, use of the one-loop triviality criteria would give the most generous range for the quartic couplings $\lambda(M_Z)$, $b_4(M_Z)$, and $a_2(M_Z)$ were it not for the incidence of unstable minima at scales beyond the perturbative regime.  A full two-loop analysis of the \model~-- together with a study of scattering amplitudes that depend on these parameters -- would lead to a precise statement of perturbativity for the quartic couplings and a more restricted range of parameter space. However, given the subtle interplay of the various new parameters that occurs already at one-loop order and the qualitatively new features that emerge, we defer a refined study of two-loop stability and perturbativity considerations to future work. 

In the interim, for the purposes of this analysis we consider two possible criteria to constrain the values of the couplings at the cutoff scale $\Lambda$, hence leading to a more trustworthy perturbative loop expansion of the potential.  The first option is to take the SM two-loop result and apply it to each of the quartic couplings at the cutoff scale individually:
\begin{equation}\label{perturbativity_less}
	\begin{aligned}
		\lambda\of{\Lambda} &< \lambda_\mathrm{FP}/4\\
		a_2\of{\Lambda} &< \lambda_\mathrm{FP}/4\\
		b_4\of{\Lambda} &< \lambda_\mathrm{FP}/4\ \ ,
	\end{aligned}
\end{equation}
where $\lambda_\mathrm{FP}/4\simeq 18$ is taken to be a fixed number (in the normalization of Ref.~\cite{Hambye:1996wb}, $\lambda_\mathrm{FP}\simeq 12$).  This is the less restrictive of the two perturbative criteria we impose.  The second, more restrictive choice echoes the constraints imposed by the authors of Ref.~\cite{Lerner:2009xg} (adapted to our normalizations):
\begin{equation}\label{perturbativity_more}
	\begin{aligned}
		\lambda\of{\Lambda} &< 4\pi\\
		a_2\of{\Lambda} &< 2\pi\\
		b_4\of{\Lambda} &< 2\pi/3\ \ .
	\end{aligned}
\end{equation}
From a numerical study of the RG-improved one-loop effective potential, we find that using either of these two options allows us to avoid the unstable minima in the potential that arise when the triviality criterion is used.  Additionally, these perturbativity restrictions indicate a range where the upper bounds of a more extensive two-loop analysis may lie.

With these considerations in mind, we translate the bounds on $\lambda(M_Z)$ into bounds on $M_h$ by working at the scale $\mu=v_0$ where we impose the minimization condition in Eq.~\eqnref{One_loop_minimization_higgs}. The corresponding eigenvalues of the mass-squared matrix are given by
\bea
m_h^2\ofv &=& \frac{\partial^2\veff}{{\partial h}^2}\Biggr\vert_{h=v_0,\,S=0,\,\mu=v_0}
	\label{One_loop_higgs_running_mass}\\
m_S^2\ofv &=& \frac{\partial^2\veff}{{\partial S}^2}\Biggr\vert_{h=v_0,\,S=0,\,\mu=v_0}
\label{One_loop_singlet_running_mass}\ \ \ .
\eea
The unbroken $Z_2$ symmetry of the potential implies that the off-diagonal term
\be
m_{hS}^2 = \frac{\partial^2\veff}{\partial h\partial S} \Biggr\vert_{h=v_0,\,S=0,\, \mu=v_0}=0 
\ee
so that the doublet and singlet fields do not mix. 

The physical masses of the Higgs and singlet, $M_h^2$ and $M_S^2$, are determined by the poles of the propagators.  In the $\msbar$ scheme, they are related to the running masses ${\hat m}_h^2(\mu)$ and ${\hat m}_S^2(\mu)$ by
solving 
\begin{align}
	&0 = M_h^2-\lp {\hat m}_h^2(\mu) + \widehat\Sigma_h\lp p^2=M_h^2, \mu \rp\rp\label{Higgs_pole_mass}\\
	&0 = M_S^2-\lp  {\hat m}_S^2(\mu) + \widehat\Sigma_S\lp p^2=M_S^2, \mu\rp\rp\label{Singlet_pole_mass}\ \ \ ,
\end{align}
where $\widehat\Sigma_{h,S}$ denote the $\msbar$ renormalized self energy functions. The eigenvalues $m_h^2$ and $m_S^2$ appearing in Eqs.~\eqnref{One_loop_higgs_running_mass} and \eqnref{One_loop_singlet_running_mass}, however, include contributions to the self-energy functions evaluated at $p^2=0$, since the one-loop effective potential contains the sum of all one-particle irreducible one-loop graphs with zero external momentum. Consequently, when relating $m^2_{h,S}$ to the pole masses $M^2_{h,S}$ we must subtract out the $p^2=0$ self energy contributions as discussed in Ref.~\cite{Casas:1994us}:
\begin{align}
	&0 = M_h^2-\lp {m}_h^2+ \Delta\widehat\Sigma_h\lp p^2=M_h^2, \mu \rp\rp\label{Higgs_corrected_pole_mass}\\
	&0 = M_S^2-\lp  {m}_S^2 + \Delta\widehat\Sigma_S\lp p^2=M_S^2, \mu\rp\rp\label{Singlet_corrected_pole_mass}\ \ \ ,
\end{align}
where 
\be
\Delta\widehat\Sigma_i \equiv \widehat\Sigma_i\lp p^2=M_i^2\rp - \widehat\Sigma_i\lp p^2=0\rp\label{Definition_of_delta_sigma}\ \ \ .
\ee

The self-energy correction to the Higgs $\Delta\widehat\Sigma_h$ is calculated in the Standard Model in the appendix of Ref.~\cite{Casas:1994us} (the authors use the notation $\Delta\Pi$), but there is an additional contribution to the self-energy from the scalar singlet which must be calculated.  The singlet contributes only one diagram to the Higgs self-energy, and the result is
\begin{equation}\label{Scalar_singlet_higgs_self_energy_correction}
	\Delta\widehat\Sigma_h^{\lp S\rp} = \frac{1}{8\pi^2}a_2\ofv^2 v_0^2\lb \mathcal{Z}\lp\frac{M_S^2}{M_h^2}\rp - 2\rb
\end{equation}
where the function $\mathcal{Z}$ is defined in Ref.~\cite{Casas:1994us} as
\begin{align}
	\mathcal{Z}\lp x\rp &= 2+\int_0^1 dy\log{\lb 1-\frac{y}{x}\lp 1-y\rp-i\epsilon\rb}\nonumber\\
	&= \left\{\begin{aligned}
			&2\sqrt{\left|1-4x\right|}\arctan{\lp 1/\sqrt{\left|1-4x\right|}\rp} & &\text{for}\,x>\frac{1}{4}\\
			&\sqrt{\left|1-4x\right|}\log{\lb\lp 1+\sqrt{\left|1-4x\right|}\rp/\lp 1-\sqrt{\left|1-4x\right|}\rp\rb} 
				& &\text{for}\,x<\frac{1}{4}
		\end{aligned}\right.\nonumber
\end{align}
On the other hand, there is only one diagram which contributes to $\Delta\widehat\Sigma_S$.  This diagram gives the result
\begin{align}
	\Delta\widehat\Sigma_S &= \frac{1}{4\pi^2}a_2\ofv^2 v_0^2\int_0^1 dx
		\log{\frac{M_h^2\lp 1-x\rp + M_S^2x^2}{M_h^2\lp 1-x\rp + M_S^2x}}\label{Singlet_self_energy_correction}\\
	&= \frac{1}{4\pi^2}a_2\ofv^2 v_0^2\lb \mathcal{Y}\lp\frac{M_S^2}{M_h^2}\rp + \lp\frac{M_h^2}{M_h^2-M_S^2}
		-\frac{M_h^2}{2M_S^2}\rp\log{\frac{M_S^2}{M_h^2}} - 1\rb\nonumber
\end{align}
where the new function $\mathcal{Y}$ is defined as
\begin{equation}\label{Special_Y}
	\mathcal{Y}\lp x\rp =\left\{\begin{aligned}
		&\frac{1}{x}\sqrt{4x-1}\lb\mathrm{arccot}\lp\sqrt{4x-1}\rp+\arctan\lp\frac{2x-1}{\sqrt{4x-1}}\rp\rb & &\text{for}\,x>\frac{1}{4}\\
		&-\frac{1}{x}\sqrt{1-4x}\,\mathrm{arctanh}\lp\sqrt{1-4x}\rp & &\text{for}\,x<\frac{1}{4}
	\end{aligned}\right.
\end{equation}
The physical masses can now be calculated from Eqs.~\eqnref{Higgs_corrected_pole_mass} and \eqnref{Singlet_corrected_pole_mass}.

\section{Results \& Discussion}
\label{sec:disc}

From the expression for $\beta_\lambda$ in Eq.~\eqnref{Lambda_beta_function}, we observe that the presence of the quartic $H^\dag H S^2$ operator generates a positive contribution to the running of $\lambda$. One would thus expect that the onset of a turnover in $\veff$ along the $h$ direction to occur at larger scales than in the Standard Model, while the occurrence of the perturbativity bound would occur at relatively lower scales. As a result, both the vacuum stability and triviality bounds on $M_h$ in the \model~should be lowered for a given $\Lambda$, relative to those in the SM. This effect was demonstrated in Refs.~\cite{Clark:2009dc,Lerner:2009xg} for $\Lambda \simeq M_{Pl}$.  In what follows, we verify this expectation explicitly for cutoffs from $1~\text{TeV}$ to $10^{19}~\text{GeV}\simeq M_{Pl}$ and for suitable choices of the boundary conditions on $b_2$ and $b_4$ leading to minimal impact of these parameters on the vacuum structure of the theory.

In general, however, the singlet quadratic and quartic parameters can play a non-trivial role.  To see why, consider Eq.~(\ref{eq:mstree}). Depending on the value chosen for $a_2$, a negative value for $b_2$ at the input scale may be required to obtain a desired value of $m_S^2$. A negative value for $b_2\of{\mu_0}$ may lead to an additional minimum for $S\not=0$ that will be deeper than the electroweak minimum unless $b_4$ is sufficiently large. These considerations are particularly important for a scalar dark matter candidate that is relatively light. The coupling $a_2$ controls the annihilation cross section, and it must be sufficiently large to prevent a relic density that is too large. On the other hand, the corresponding value of the light mass may require a negative $b_2$. In addition, larger values of $b_4$ at the input scale can cause both $b_4$ and $a_2$ to reach their perturbativity bounds sooner, modifying the perturbativity upper bounds based solely on the running of $\lambda$. Below, we analyze these effects in detail\footnote{Note that we only consider positive values of the coupling $a_2$.  In principle, a negative coupling is allowed, in which case stability of the electroweak minimum imposes a tree level constraint on $a_2$.  When including quantum corrections and one-loop running, Eq.~\eqnref{eq:a2_beta_function} indicates that a negative value at the input scale would cause $a_2$ to become even more negative very quickly with increasing scale (due largely to the top quark contribution).  This would cause the effective potential to be unbounded below in some direction of field space between the $h$ and $S$ axes.  Requiring stability of the potential up to a cutoff scale $\Lambda$ would then impose a lower bound on $a_2$.  A negative $a_2$ coupling can contribute to a strong first-order electroweak phase transition \cite{Profumo:2007wc}, but does not affect the dark matter phenomenology because it enters the equations for the annihilation and direct detection cross sections as $a_2^2$.}.

\begin{figure}
	\includegraphics{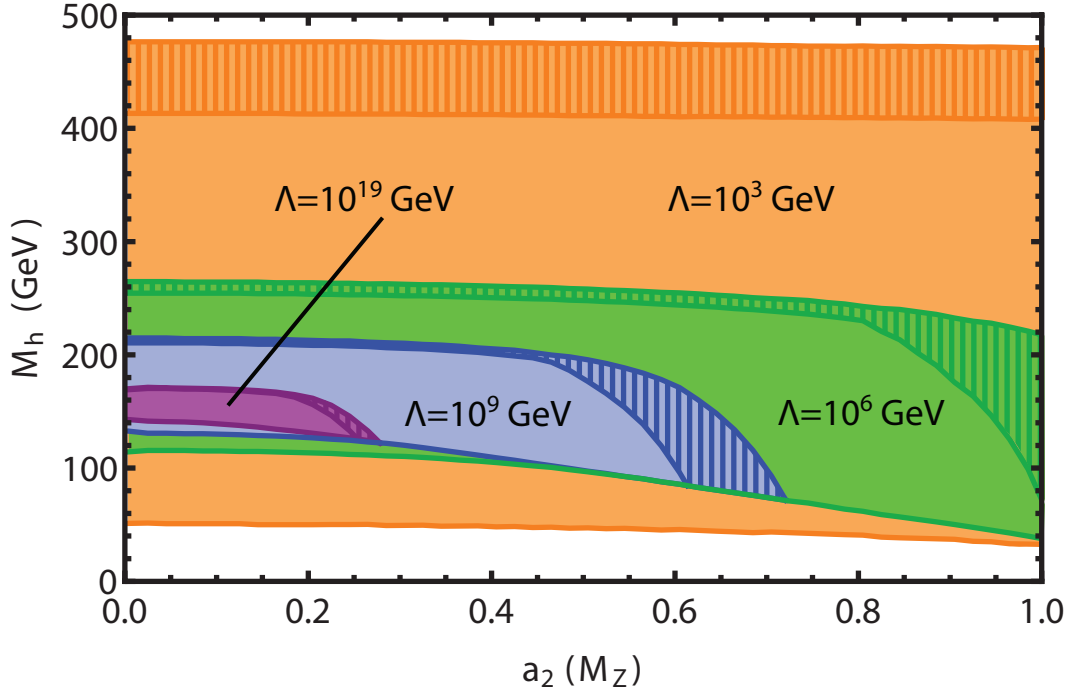}
	\caption{The change in the Higgs mass upper and lower bounds as $a_2\of{M_Z}$ is increased while requiring stability of the potential and perturbativity of the quartic coupling constants up to cutoff scales $\Lambda=\of{10^{19}\comma 10^9\comma 10^6\comma 10^3}\unit{GeV}$ (the purple [dark gray], blue [light gray], green [medium gray], and orange [light gray] regions, respectively).  Here, $b_4\of{M_Z}=0.001$ and $b_2\ofv = 0$.  Perturbativity of the potential using the constraints of Eq.~\eqnref{perturbativity_more} is indicated by the solid-colored regions; the combined solid-colored and striped regions indicate perturbativity of the potential using Eq.~\eqnref{perturbativity_less}.}
	\label{fig:mhva2}
\end{figure}

\begin{figure}
	\includegraphics{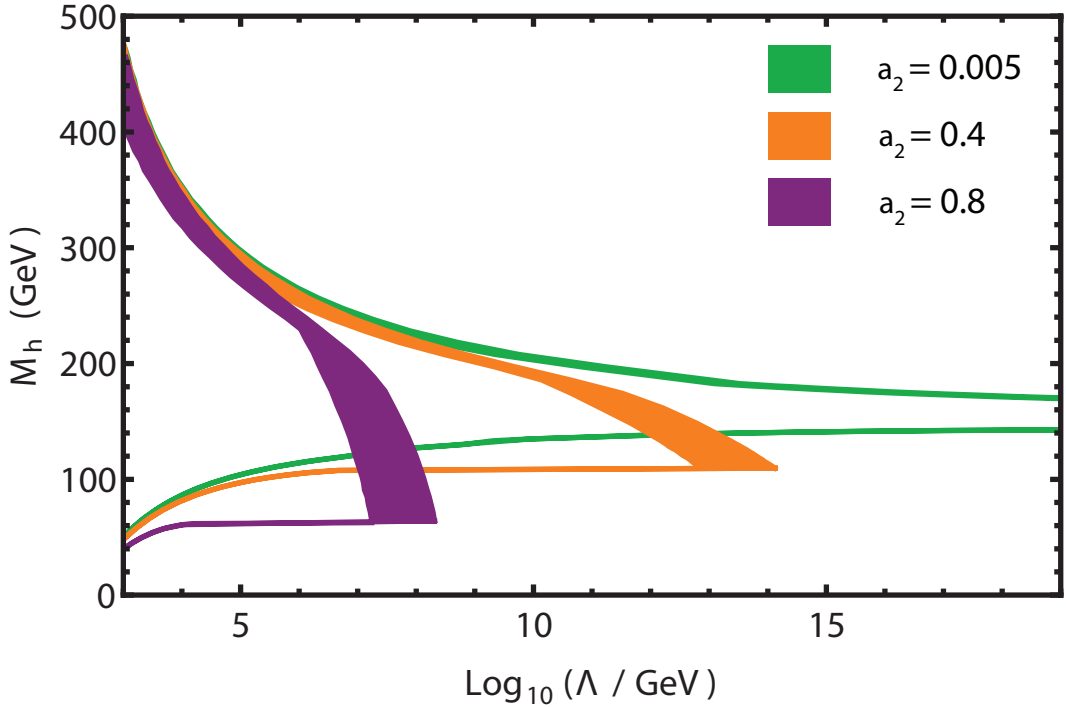}
	\caption{One-loop vacuum stability and perturbativity bounds on the Standard Model Higgs mass $M_h$ as a function of the cutoff $\Lambda$ for a set of values of $a_2\of{M_Z}$.  We show $a_2\of{M_Z}=\of{0.005, 0.4, 0.8}$ using green [medium gray], orange [light gray], and purple [dark gray], respectively.  The upper bounds show the variation between the less restrictive constraints of Eq.~\eqnref{perturbativity_less} and the more restrictive constraints of Eq.~\eqnref{perturbativity_more}.}
	\label{fig:funnelall}
\end{figure}

We first illustrate the impact of the new interactions associated with the singlet scalar on the stability of the electroweak minimum and the occurrence of the perturbative limits in the Higgs quartic self-coupling. In Fig.~\ref{fig:mhva2} we show the range of allowed Higgs masses as a function of its coupling to the singlet, $a_2$, while requiring stability of the potential up to $10^{19},\ 10^9,\ 10^6,\ \text{or}\ 10^3~\text{GeV}$.  In Fig.~\ref{fig:funnelall} we plot $M_h$ against the cutoff scale for representative values of $a_2$. For Figs.~\ref{fig:mhva2} and \ref{fig:funnelall}, we have taken the somewhat artificial choice $b_2(v_0)=0$, $b_4(M_Z)=10^{-3}$ to isolate the dependence of the Higgs mass bounds on the $a_2$ parameter. In Fig.~\ref{fig:funnelall}, the curves for successively larger values of $a_2$ illustrate the reduction in both the stability lower bound and perturbativity upper bound expected from the behavior of $\beta_\lambda$. For sufficiently large $a_2$, the impact of including this parameter in the one-loop $\beta$ function for the Higgs quartic coupling leads to a maximum $\Lambda$ (see the orange and purple curves) -- a feature that does not arise in the SM or for very small values of $a_2$ ({\em e.g.}, the green curve)\footnote{In the text, we refer to features of our figures according to their coloring.  In the captions of the figures, we provide both color and grayscale schemes.  A full color version of this work is available online.}. In short, for a sufficiently strong $H^\dag H S^2$ interaction, it is not possible to choose a value of $M_h$ that maintains both electroweak vacuum stability and perturbative behavior to arbitrarily large scales. 


The solid-colored and striped regions in Fig.~\ref{fig:mhva2} indicate the choice of perturbativity restrictions: the solid-colored regions are perturbative up to the indicated cutoff scale when the more restrictive choice of Eq.~\eqnref{perturbativity_more} is applied, while the combined solid and striped regions of each color use the less restrictive choice of Eq.~\eqnref{perturbativity_less} to maintain perturbativity to the desired $\Lambda$.  The width of the upper bands in Fig.~\ref{fig:funnelall} show the variation in the Higgs mass upper bounds when varying the perturbativity restrictions for the quartic couplings at the cutoff scale between Eqs.~\eqnref{perturbativity_more} and \eqnref{perturbativity_less}, as discussed in the previous section.  We have not shown any theoretical ranges for these bounds due to other considerations, {\em e.g.}, varying the top quark mass over its experimental range. Compared to the expected impact of including two-loop corrections and the ambiguities in choice of perturbativity bounds, the $M_t$ uncertainty is subleading. 

\begin{figure}
	\includegraphics{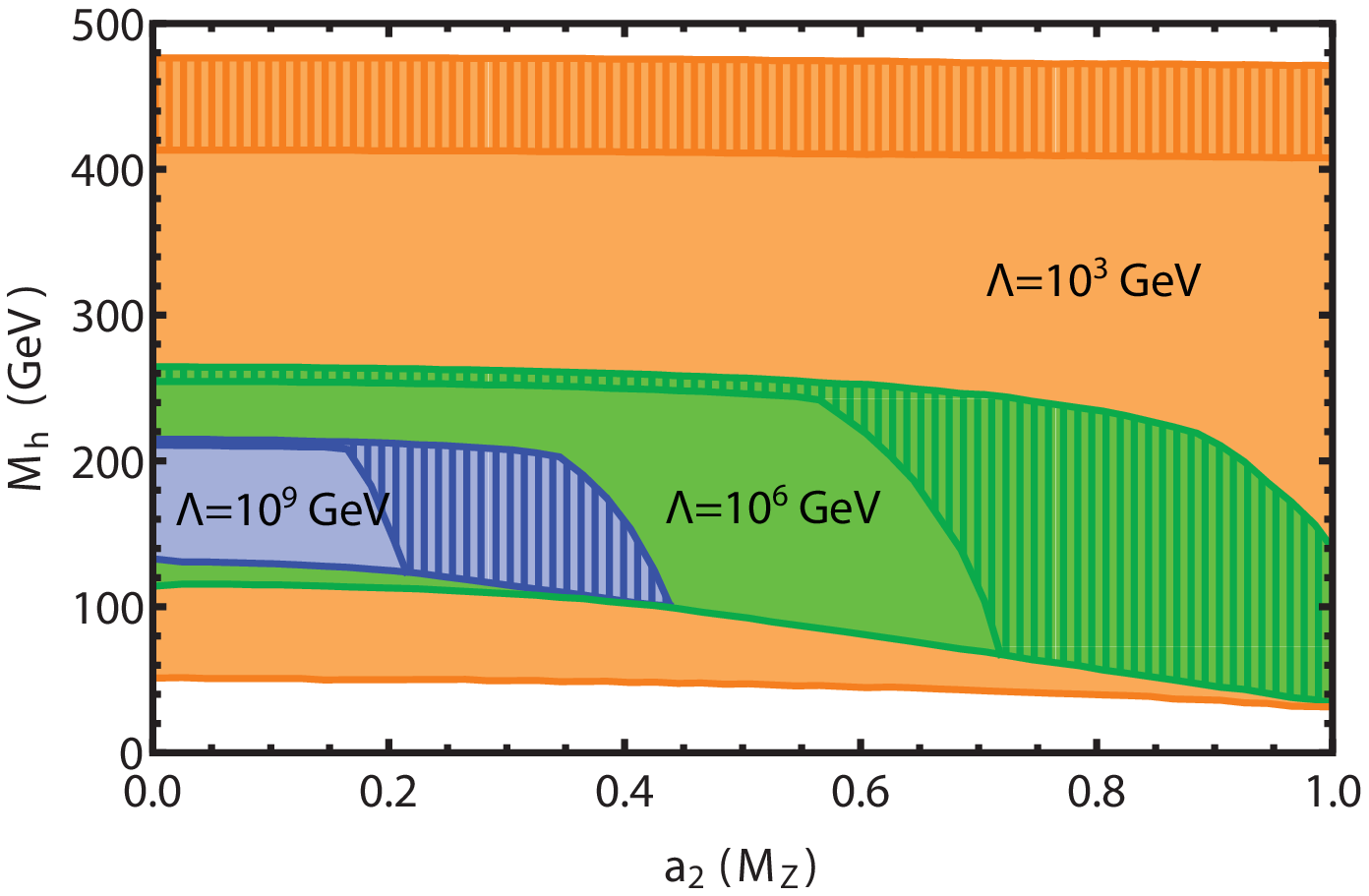}
	\caption{The analogous plot to Fig.~\ref{fig:mhva2} for the case $b_4\of{M_Z}=0.4$.  Here there is no region of Higgs masses and $S$-$H$ couplings $a_2$ that allow for perturbativity of the potential up to $10^{19}\unit{GeV}$.  The color and grayscale schemes are the same as Fig.~\ref{fig:mhva2}.}
	\label{fig:mhva2_b4}
\end{figure}

\begin{figure}
	\includegraphics{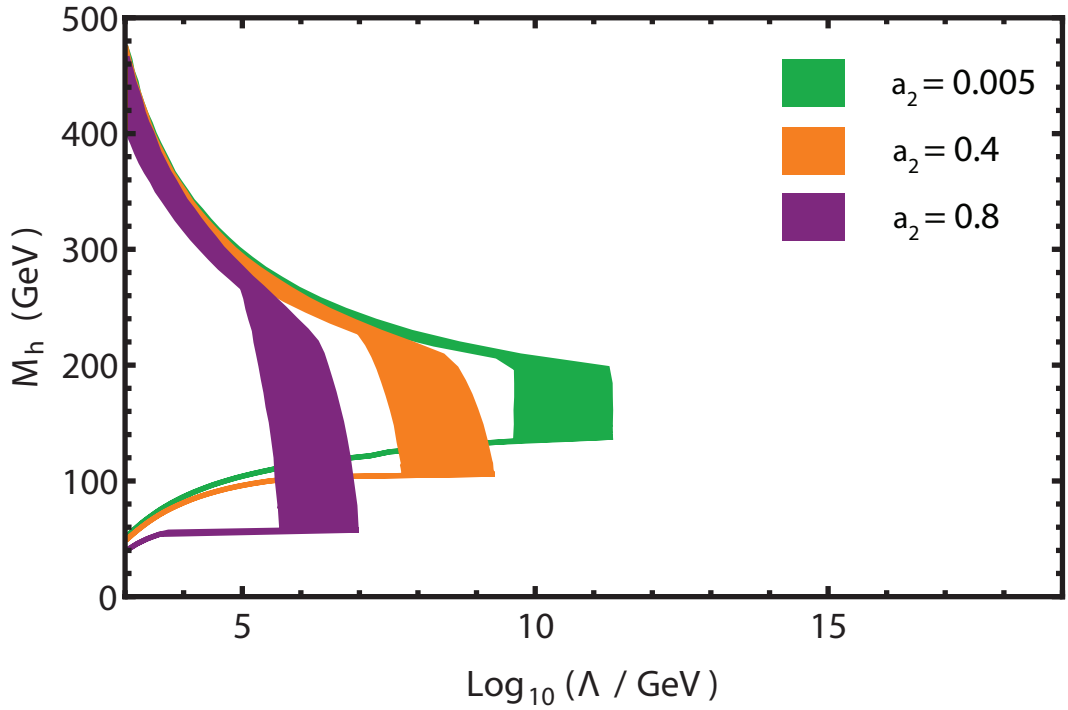}
	\caption{The analogous plot to Fig.~\ref{fig:funnelall} for the case $b_4\of{M_Z}=0.4$, demonstrating that stability and perturbativity of the potential cannot be maintained to arbitrarily large ({\em i.e.,} $\mathcal{O}\of{M_{Pl}}$) scales when $b_4$ is increased.  The color and grayscale schemes are the same as Fig.~\ref{fig:funnelall}.}
	\label{fig:funnelall_b4}
\end{figure}

In Figs.~\ref{fig:mhva2_b4} and \ref{fig:funnelall_b4}, we show analogous results to Figs.~\ref{fig:mhva2} and \ref{fig:funnelall} where $b_4\of{M_Z}$ is now taken to be $0.4$ ($b_2\ofv$ is still zero).  The lower bounds from stability of the potential in the $h$ direction are largely unchanged.  On the other hand, the upper bounds from perturbativity are reduced.  The larger value of $b_4$ can cause the couplings to reach their perturbative limits below the desired cutoff scale, and so either $\lambda\of{M_Z}$ must be reduced or it is impossible to maintain perturbativity of the couplings up to the desired cutoff.  In particular, the green curve of Fig.~\ref{fig:funnelall_b4} now demonstrates maximum scale behavior similar to the orange and purple curves, so no value of $M_h$ and $a_2$ will give a stable potential with perturbative couplings up to $M_{Pl}$.


The impact of including non-zero values for both the singlet quadratic and quartic couplings can also be significant, particularly when considering the relic density of singlet dark matter\footnote{For simplicity, we henceforth use only the perturbativity constraints in Eq.~\eqnref{perturbativity_less}.}. The $a_2$ and $M_S$ parameters control the $SS$ annihilation cross section and, thus, the singlet relic density, assuming thermal production \cite{McDonald:1993ex,Burgess:2000yq,He:2008qm,McDonald:2001vt,Silveira:1985rk,Bento:2001yk}. As discussed in Section \ref{sec:vacstab}, the region of relatively light $M_S$ and larger $a_2$ can force the parameter $b_2(v_0)$ to be negative, leading to a secondary extremum with $S\not=0$.  Thus, for some choices of the \model~couplings and masses, the extrema with $S\neq 0$ is a minimum of lower energy than the electroweak/dark matter (EW/DM) minimum corresponding to $(h=v_0\comma S=0)$; for other choices of the parameters, the extrema with $S\neq 0$ either do not exist or have greater energy than the desired EW/DM minimum.  

\begin{figure}
	\includegraphics{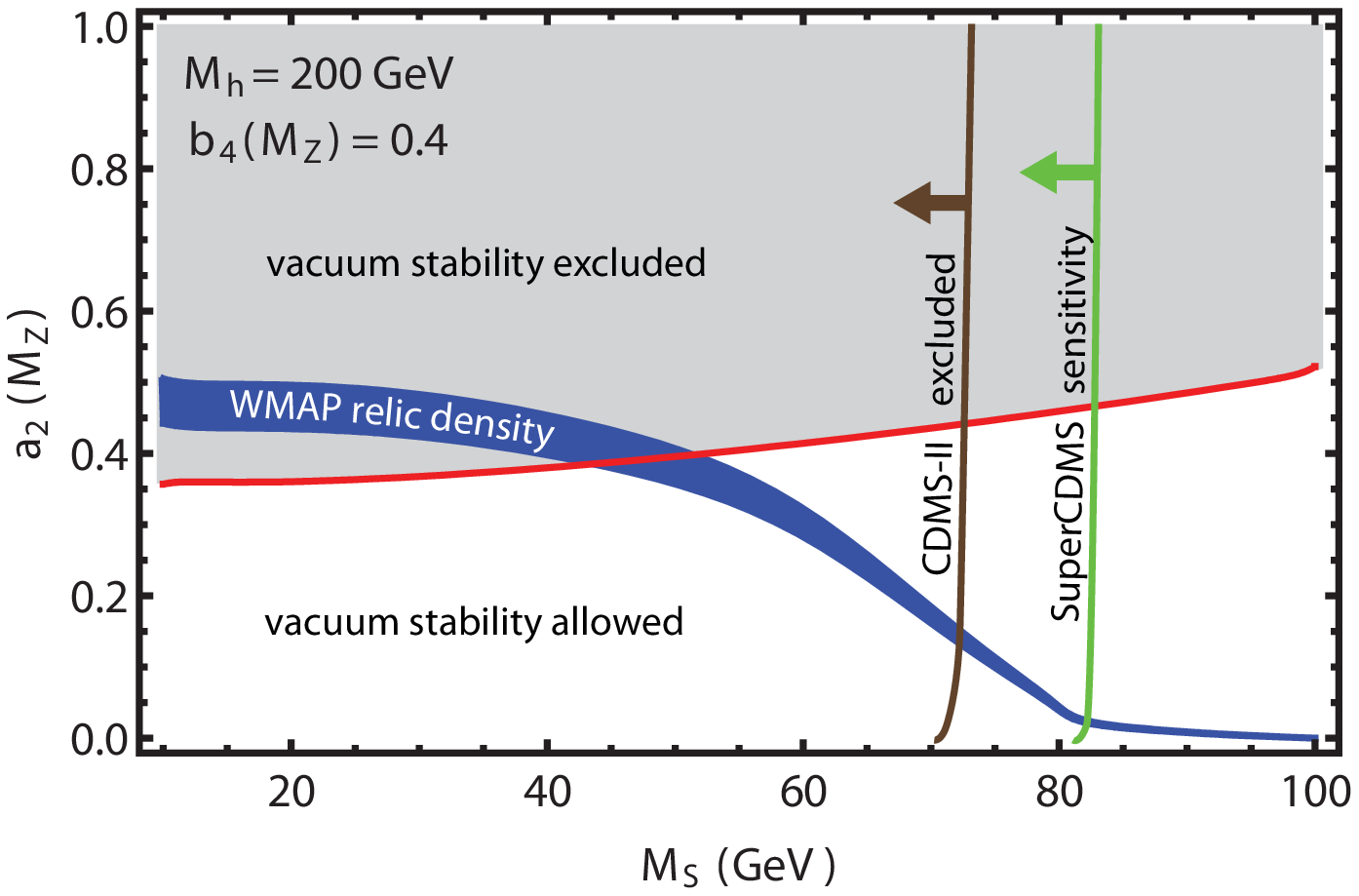}
	\caption{Values of $a_2\of{M_Z}$ and $M_S$ consistent with WMAP relic density measurements (blue [dark gray] band) and vacuum stability/perturbativity (unshaded regions) for fixed $M_h=200\unit{GeV}$ and $b_4\of{M_Z}=0.4$.  The unshaded region below the red [medium gray] curve has stable $\of{h=v_0\comma S=0}$ minima up to a $\Lambda=1\unit{TeV}$ cutoff scale.  The shaded region above the red curve is excluded because of the occurrence of deeper minima along the $S$ direction of the effective potential.  Additionally, we indicate the region excluded by CDMS-II (brown [dark gray] arrow) and the region of projected sensitivity for SuperCDMS (green [light gray] arrow).}
	\label{fig:dmplotmh200}
\end{figure}
 
To illustrate, we show in Fig.~\ref{fig:dmplotmh200} a plot of the coupling $a_2~\text{vs.}~M_S$.  We have fixed the Higgs mass, $M_h=200\unit{GeV}$, and the singlet quartic self-coupling at the input scale, $b_4\of{M_Z}=0.4$.  A cutoff scale of $\Lambda = 1\unit{TeV}$ is used.  The red curve in Fig.~\ref{fig:dmplotmh200} marks the boundary between the two regions discussed in the previous paragraph: for values of $a_2\of{M_Z}$ and $M_S$ located in the lower unshaded region, the effective potential has a stable EW/DM minimum, while $a_2$ and $M_S$ values in the shaded upper region are excluded due to the presence of unviable $S\neq 0$ minima along the $S$ direction of the potential below the cutoff scale.  This boundary is the result of our theoretical vacuum stability analysis.

\begin{figure}
	\includegraphics[width=.32\textwidth]{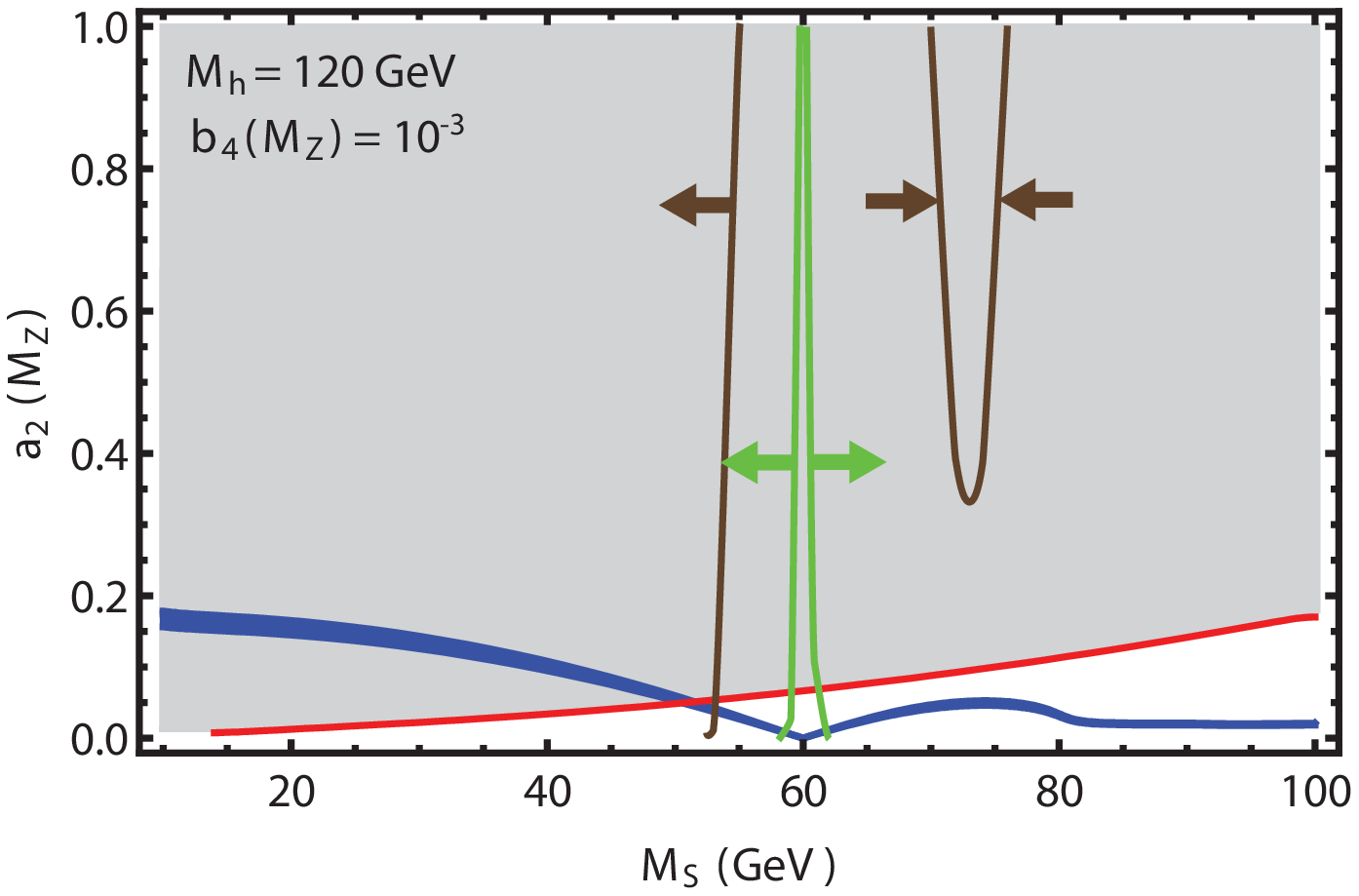}
	\includegraphics[width=.32\textwidth]{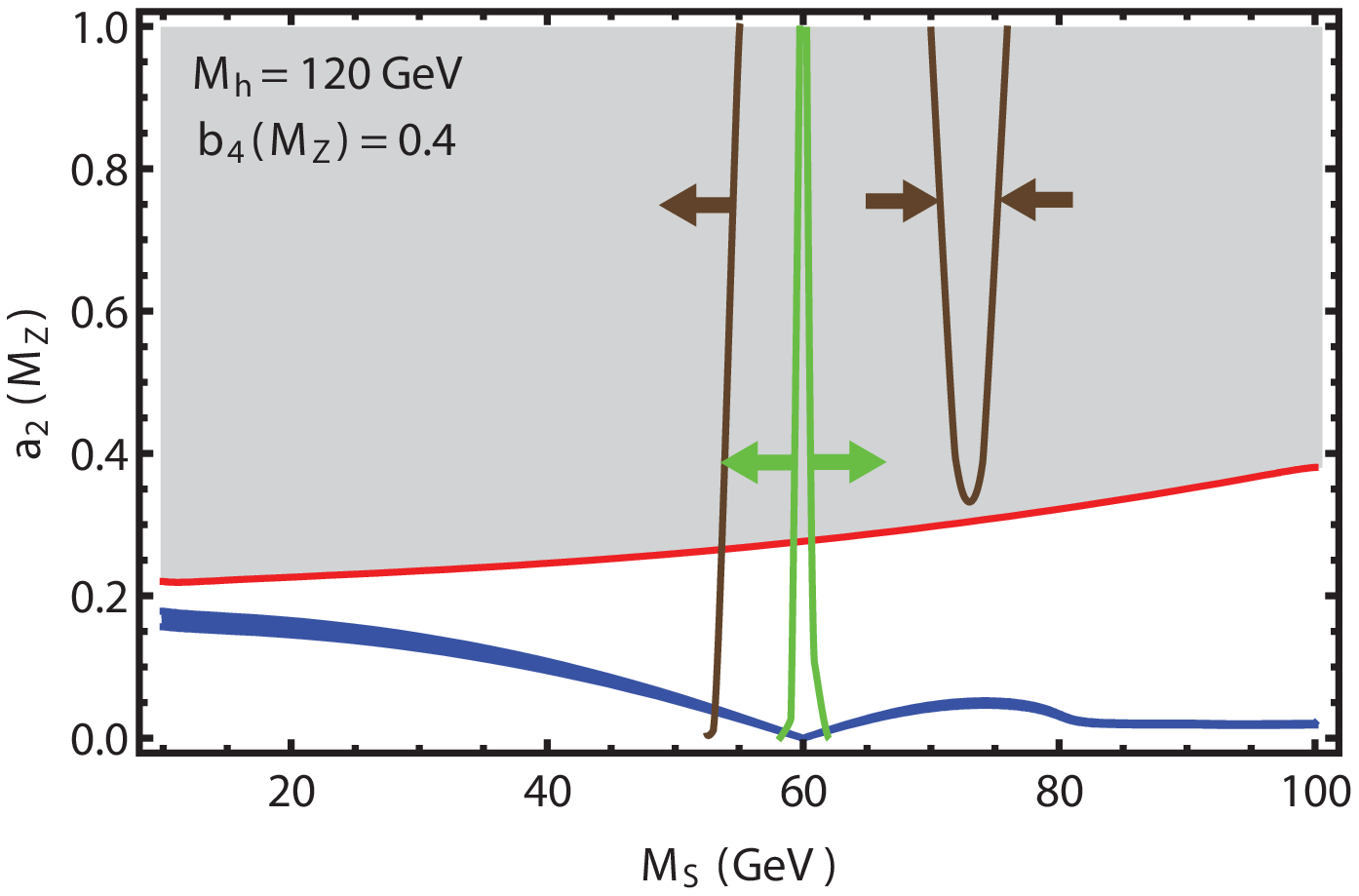}
	\includegraphics[width=.32\textwidth]{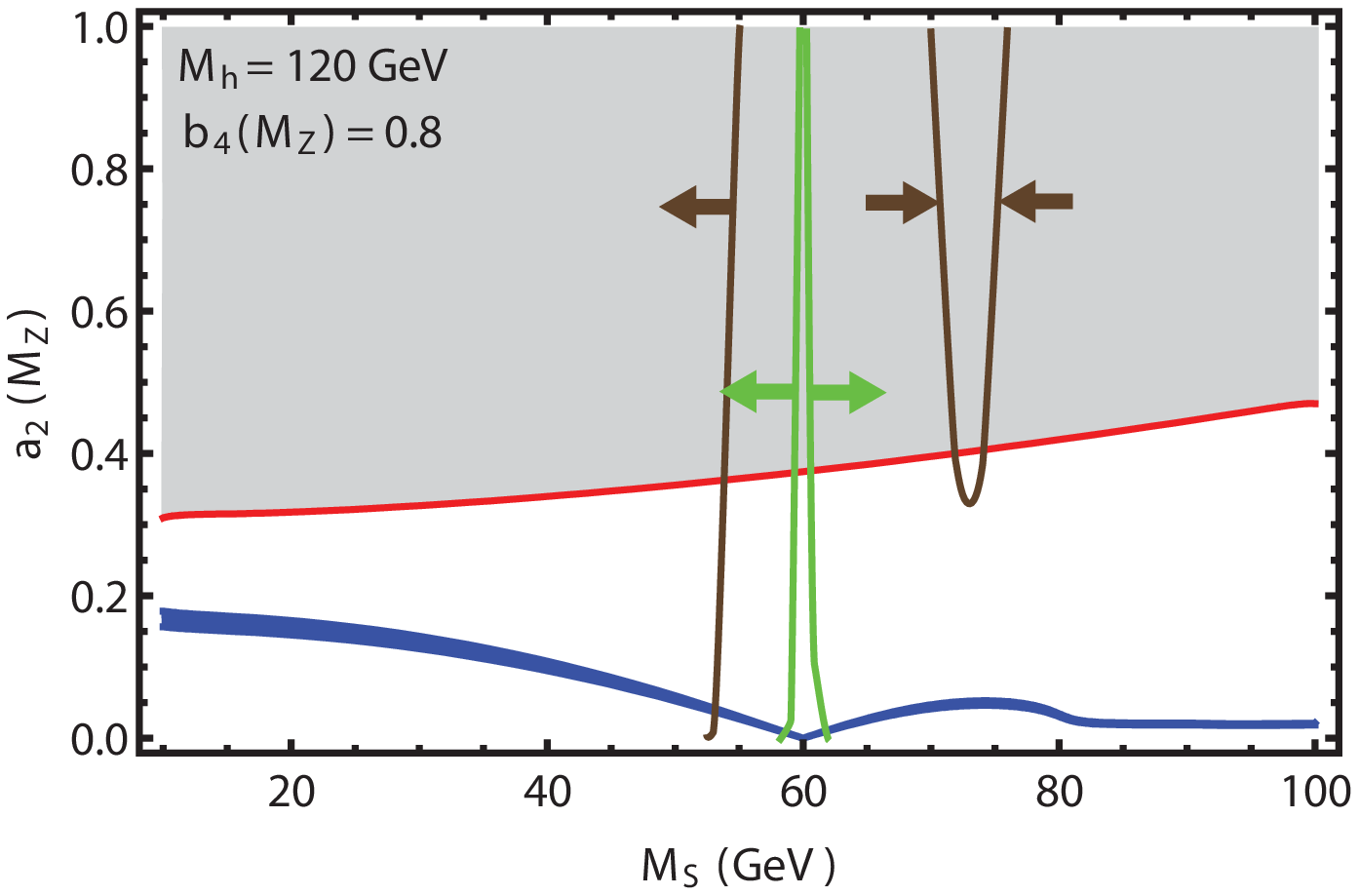}\\
	\includegraphics[width=.32\textwidth]{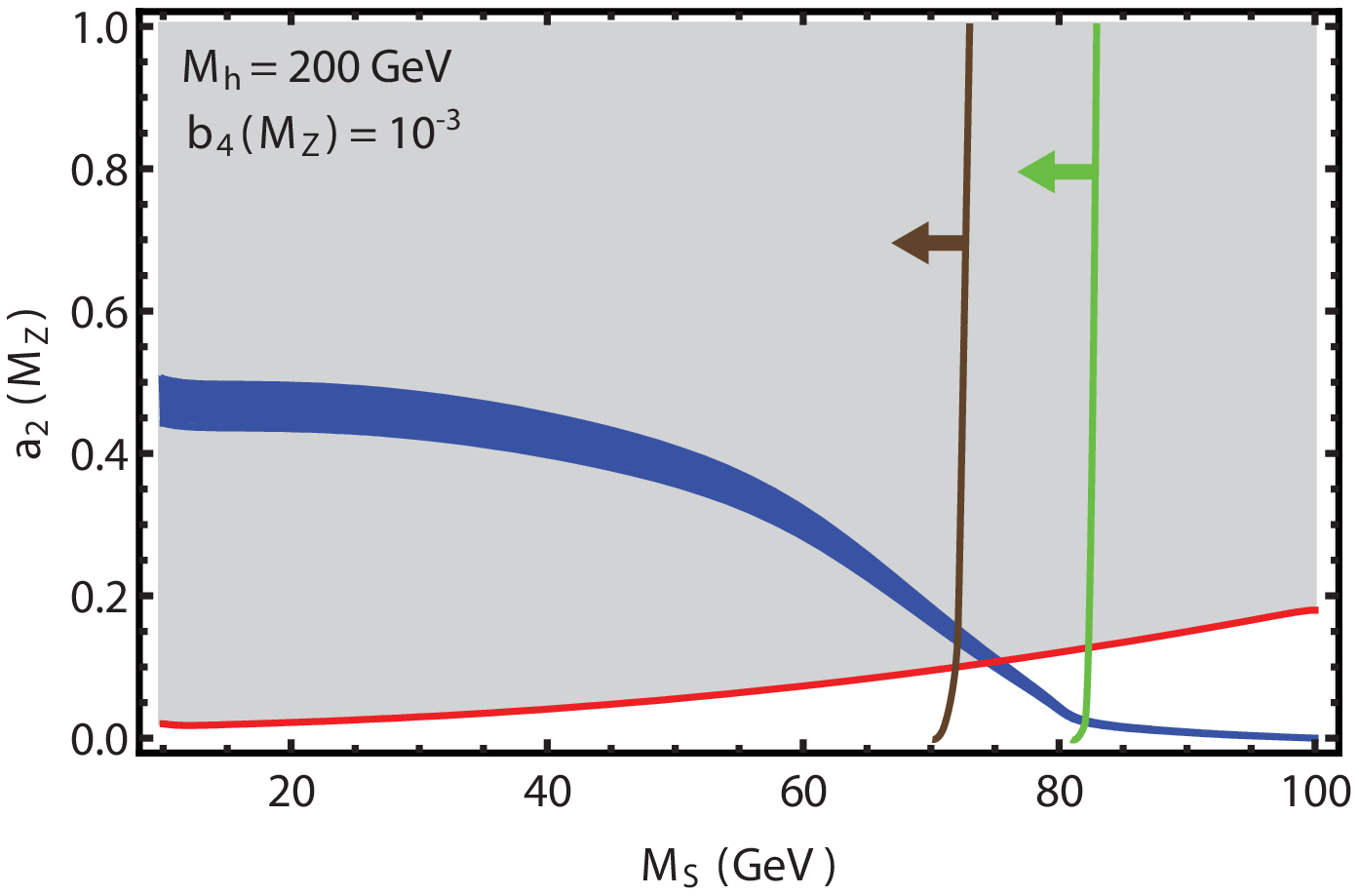}
	\includegraphics[width=.32\textwidth]{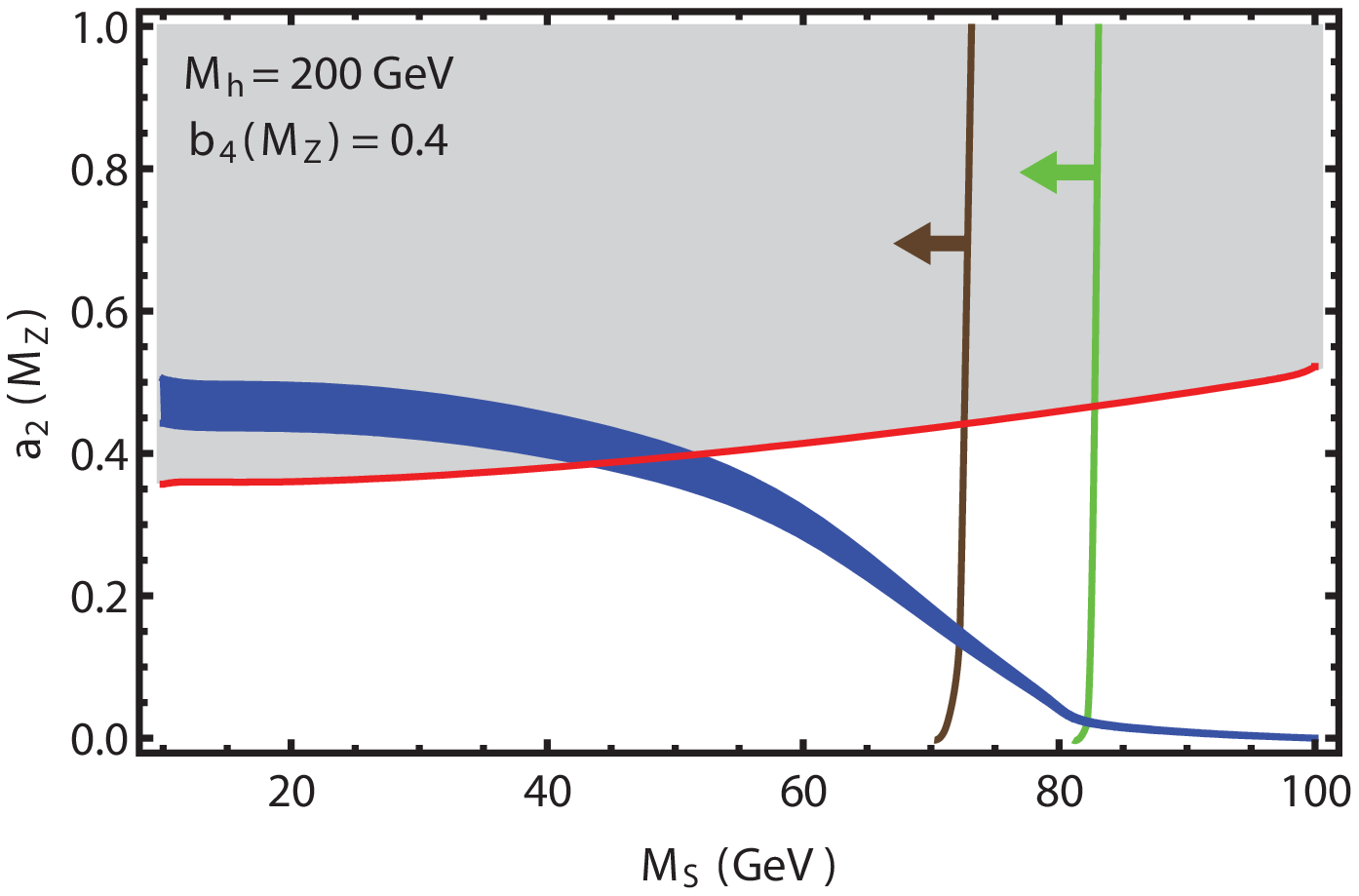}
	\includegraphics[width=.32\textwidth]{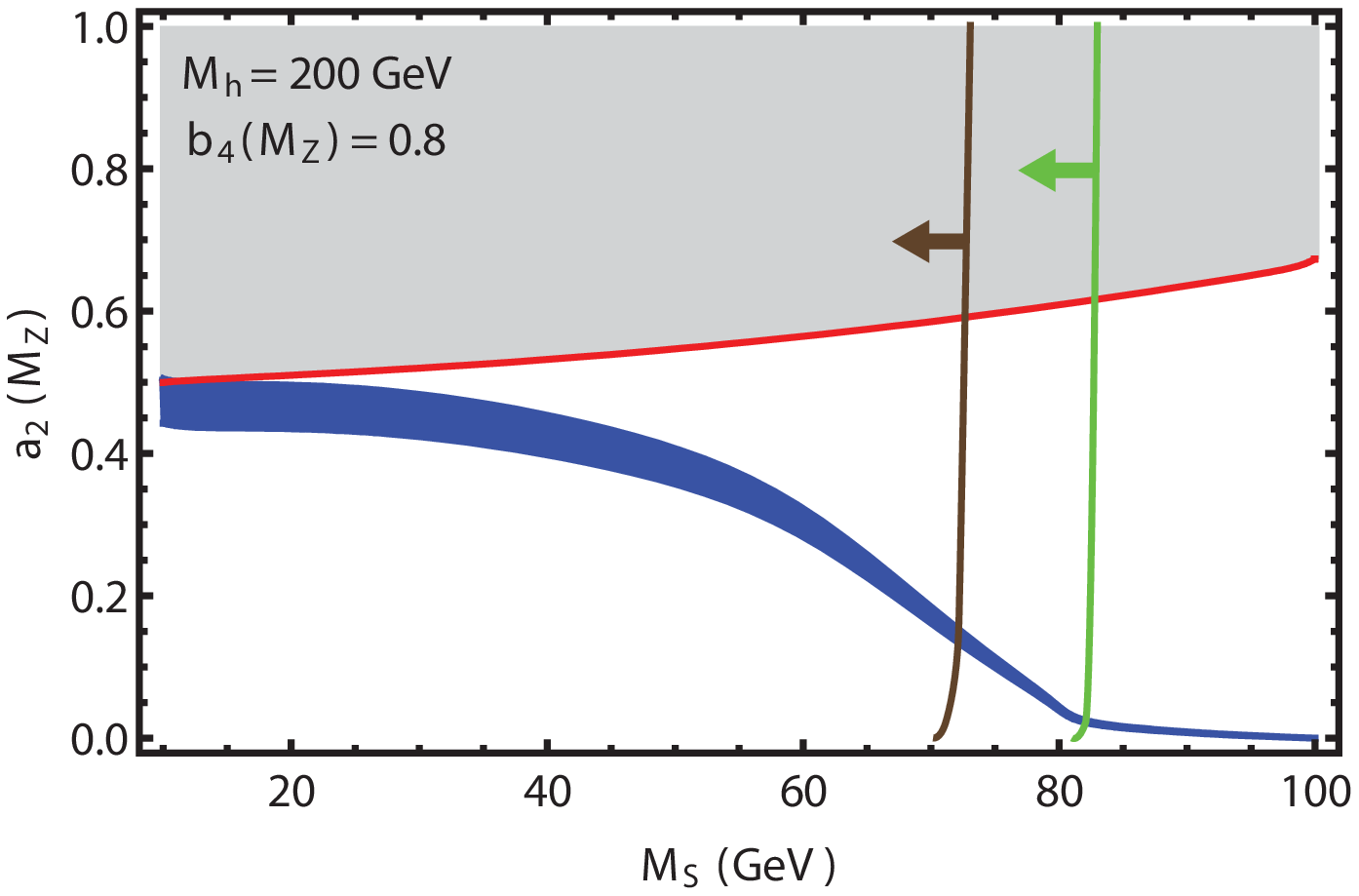}\\
	\includegraphics[width=.32\textwidth]{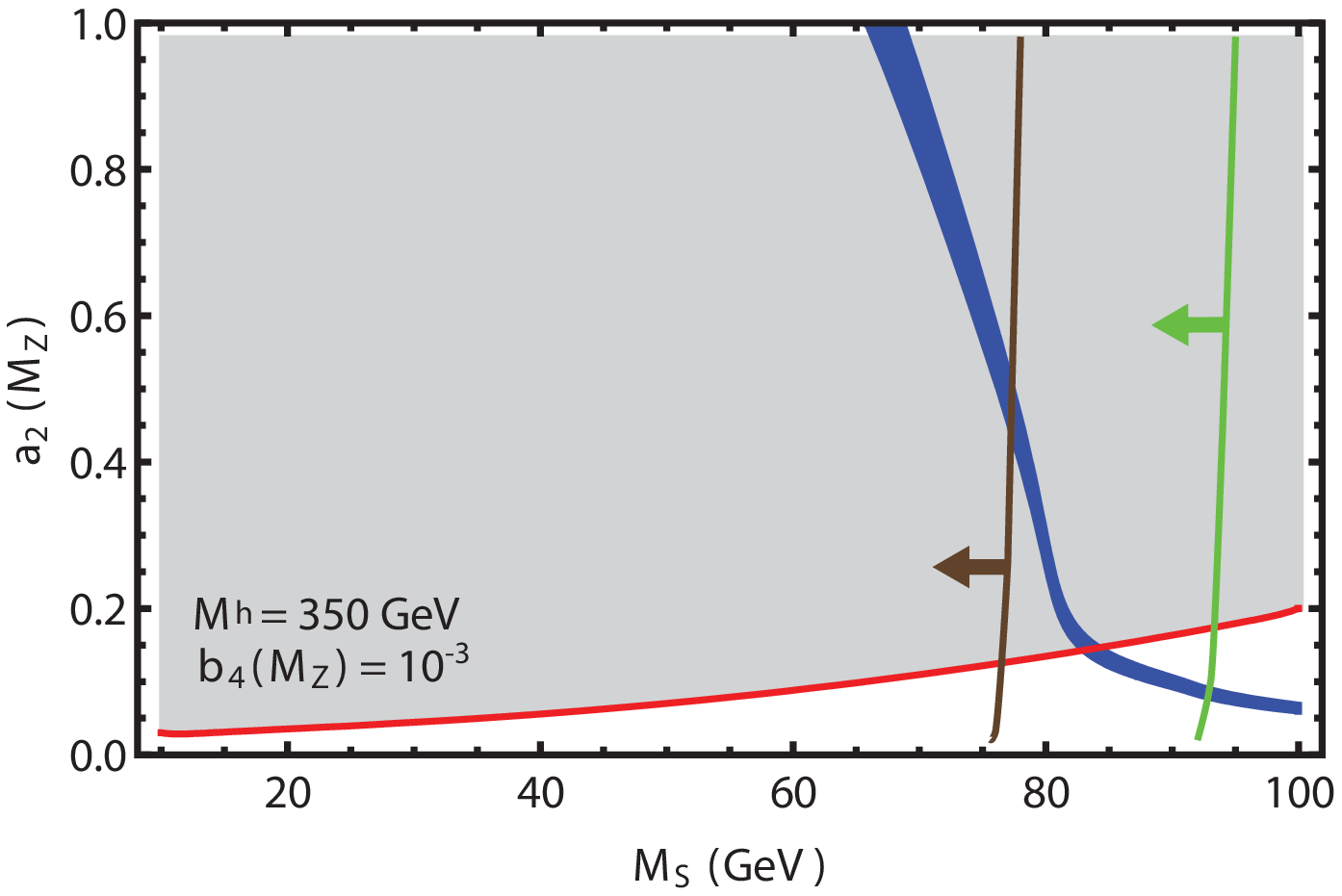}
	\includegraphics[width=.32\textwidth]{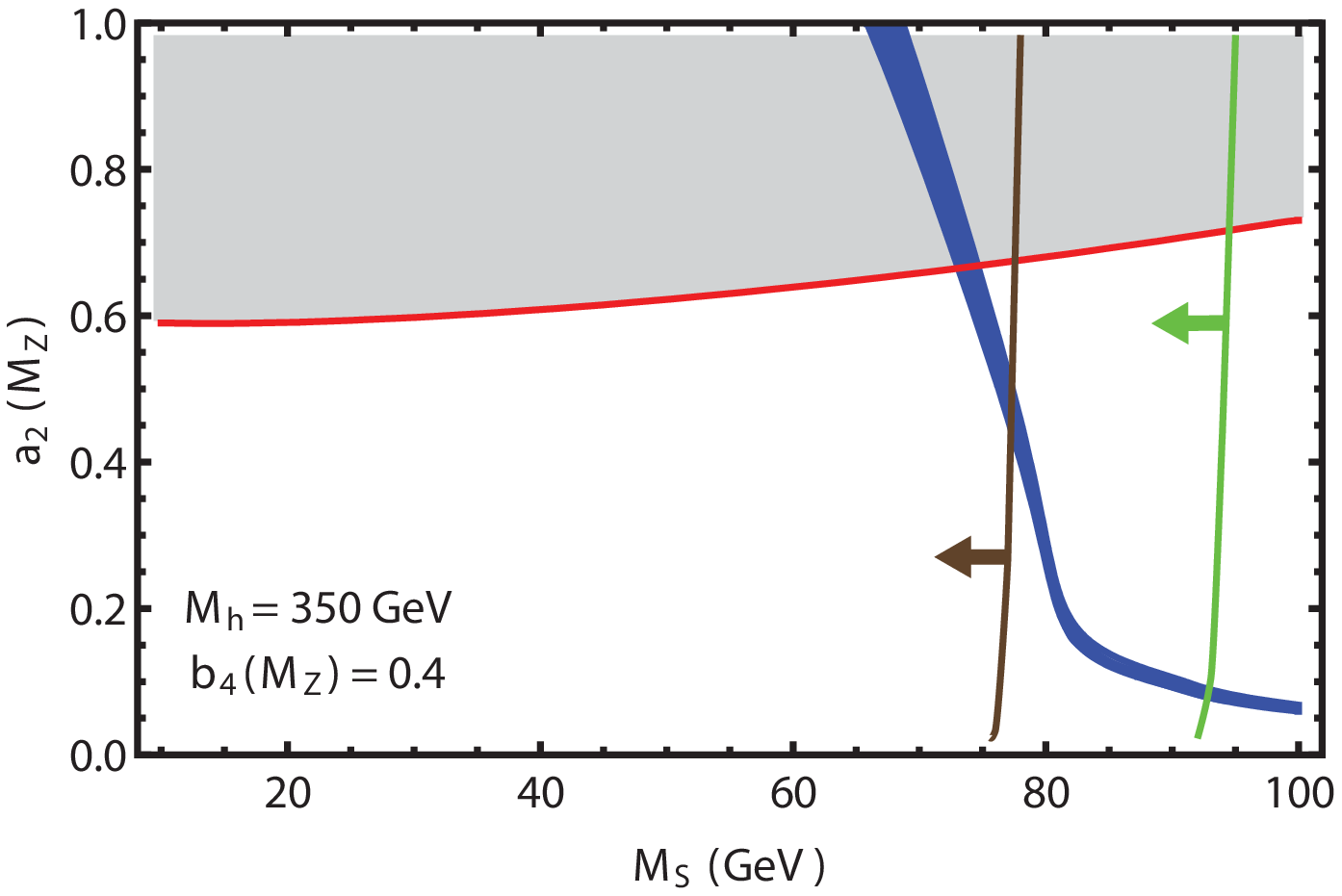}
	\includegraphics[width=.32\textwidth]{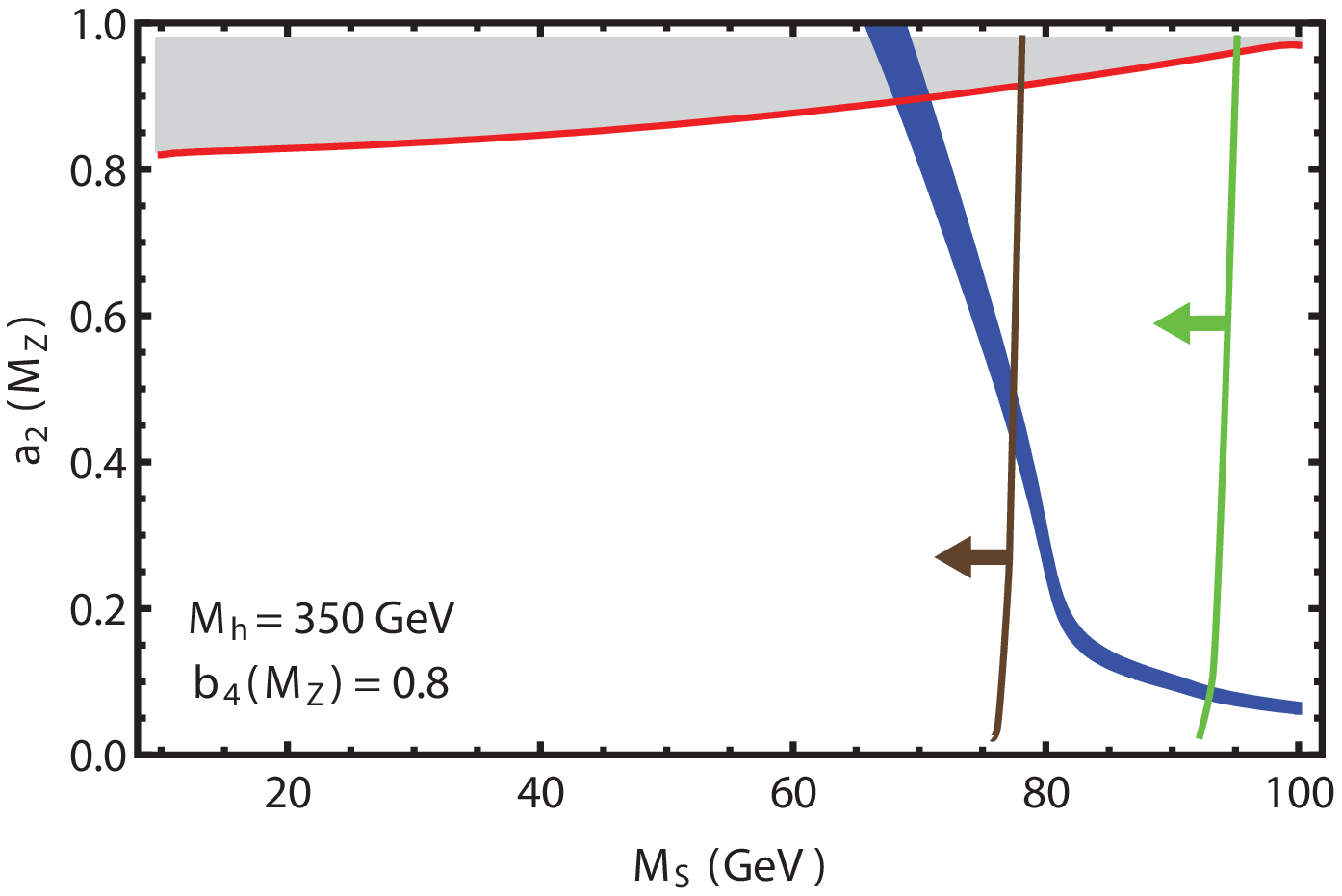}
	\caption{Values of $a_2\of{M_Z}$ and $M_S$ consistent with WMAP relic density measurements (blue [dark gray] band) and vacuum stability/perturbativity (unshaded regions).  The rows correspond to $M_h=120\unit{GeV}$ (top), $200\unit{GeV}$ (middle), and $350\unit{GeV}$ (bottom).  The columns correspond to $b_4\of{M_Z}=0.001$ (left), $0.4$ (middle), and $0.8$ (right).  Shaded regions above the red [medium gray] curves are excluded because of deeper minima along the $S$ direction.  The unshaded regions have stable EW/DM minima for a cutoff scale of $\Lambda=1\unit{TeV}$.  Additionally, we indicate the regions excluded by CDMS-II (brown [dark gray] arrows) and the regions of projected sensitivity for SuperCDMS (green [light gray] arrows).}
	\label{fig:dmplots1tev}
\end{figure}

\begin{figure}
	\includegraphics[width=.45\textwidth]{dmplot120b_5.eps}
	\includegraphics[width=.45\textwidth]{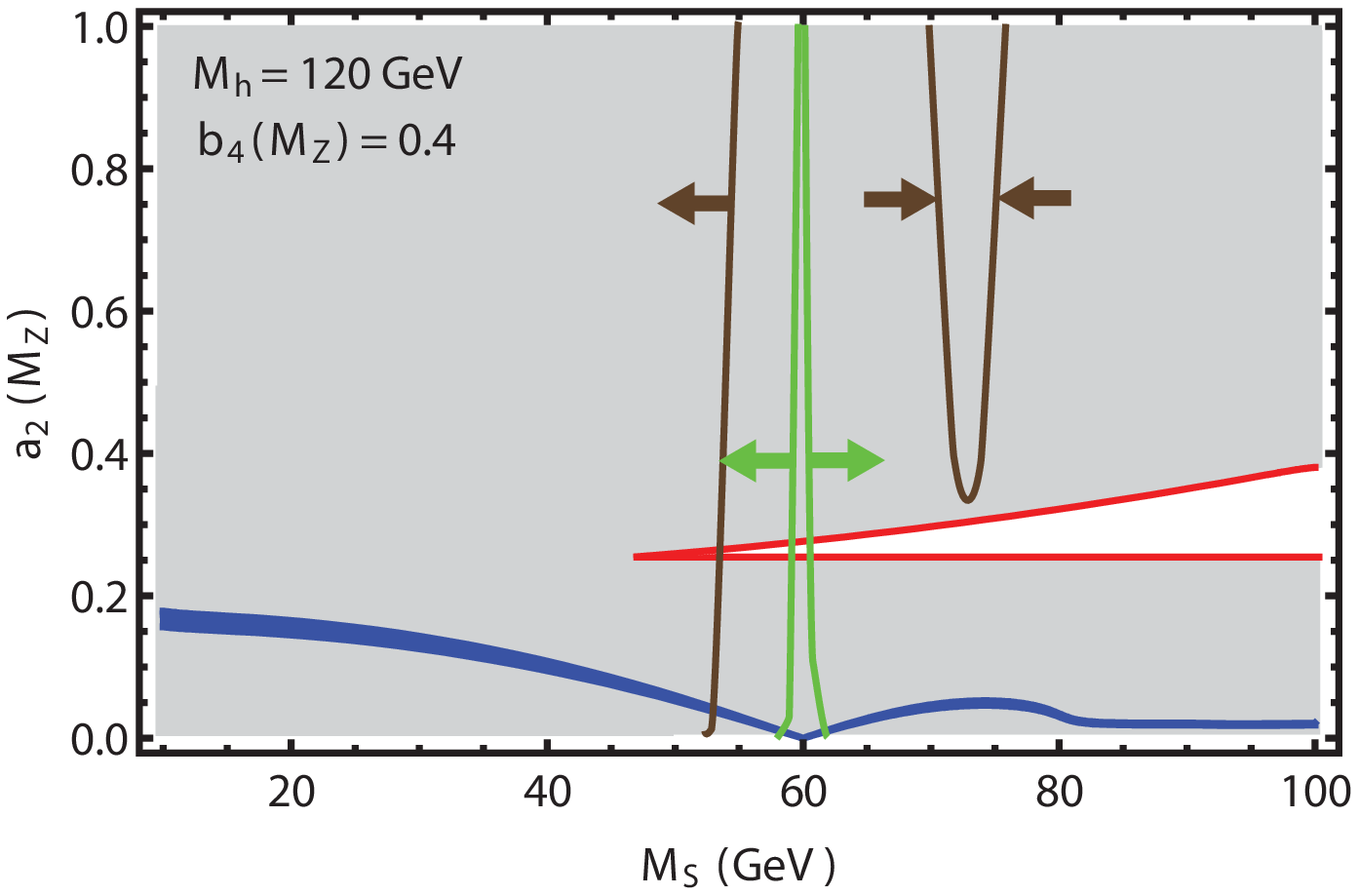}\\
	\includegraphics[width=.45\textwidth]{dmplot200b_5.eps}
	\includegraphics[width=.45\textwidth]{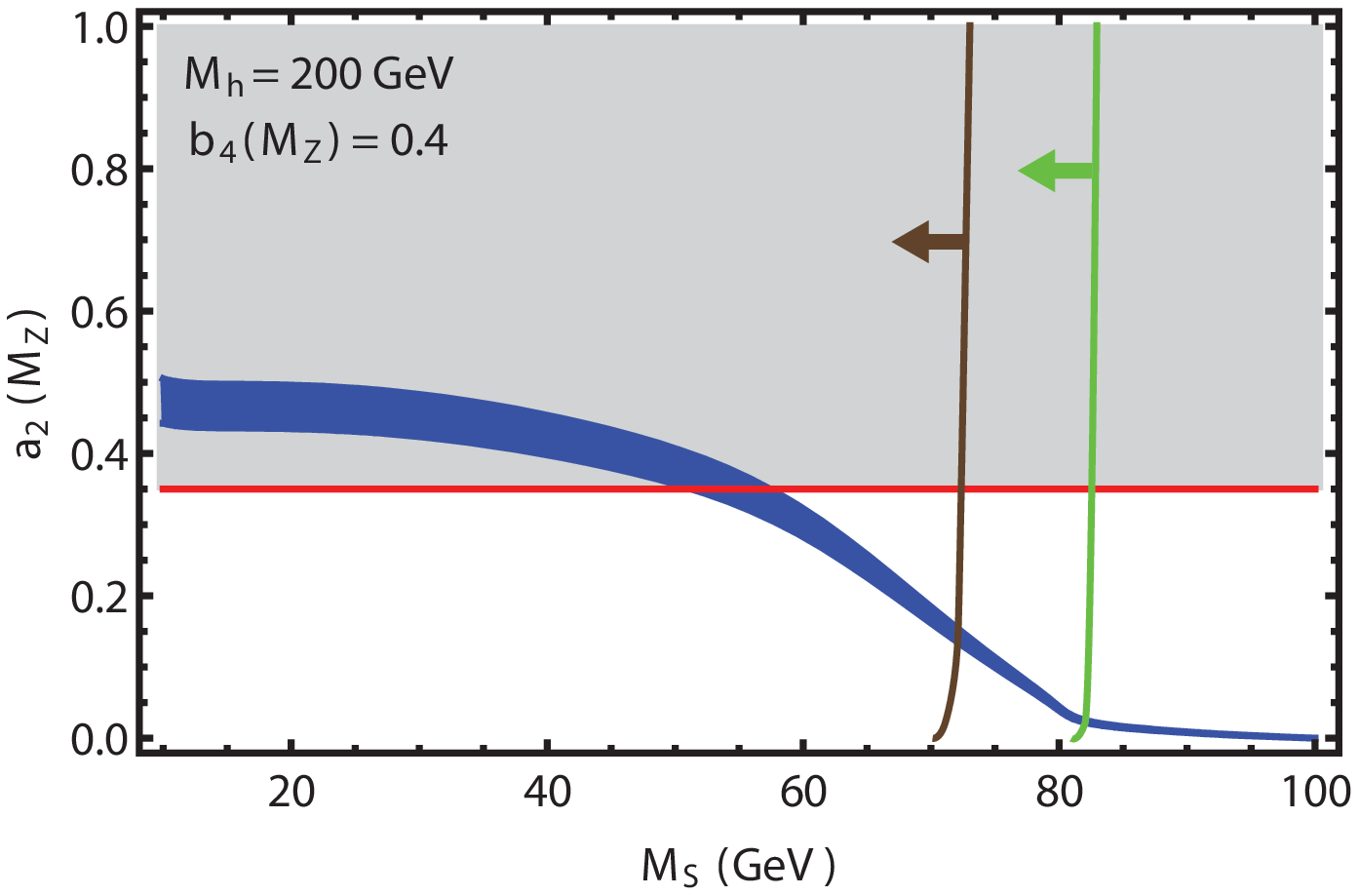}\\
	\includegraphics[width=.45\textwidth]{dmplot350b_5.eps}
	\includegraphics[width=.45\textwidth]{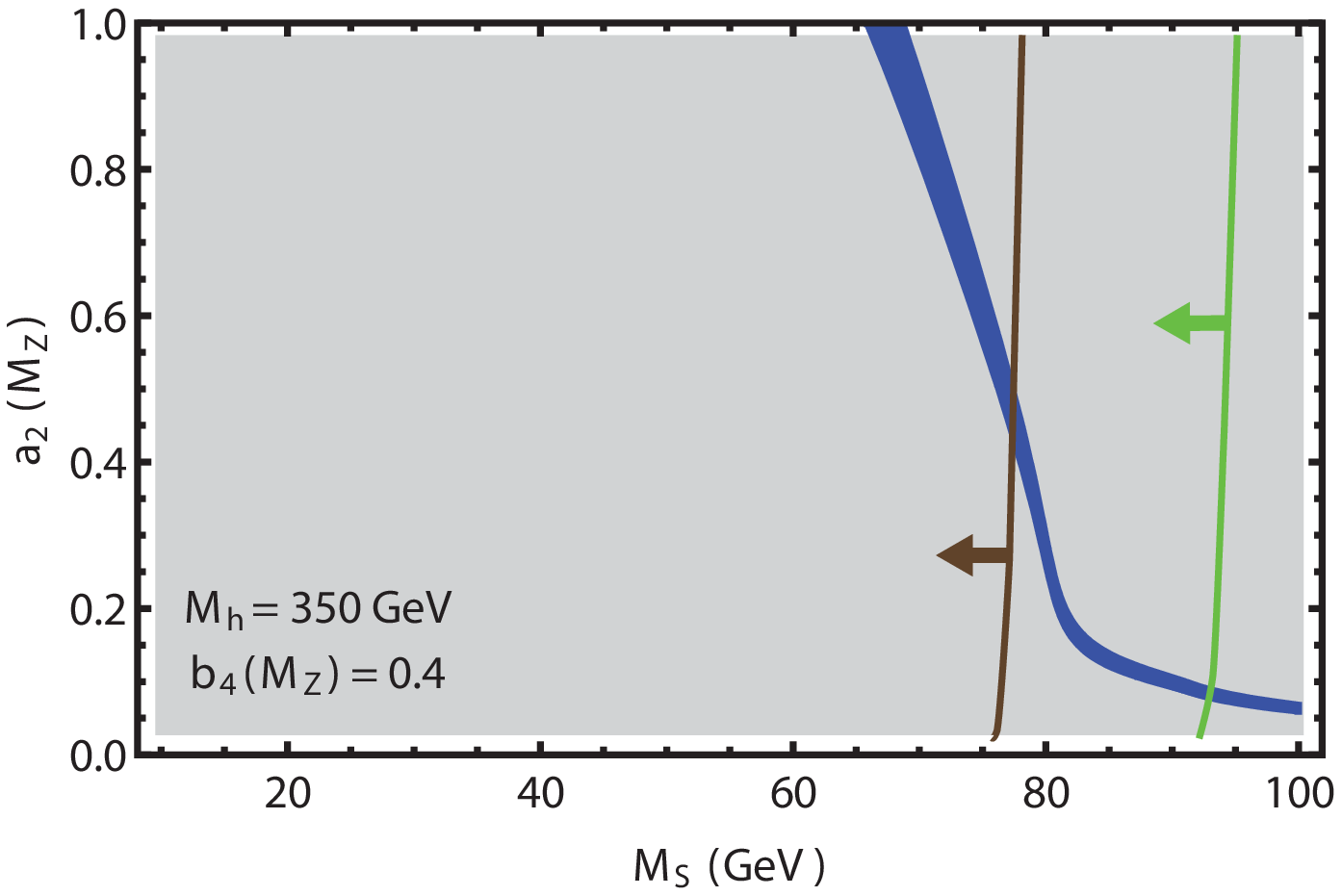}
	\caption{The column on the left is the same as the middle column of Fig.~\ref{fig:dmplots1tev} -- $M_h=120,\ 200,\ 350\unit{GeV}$ top to bottom and $b_4\of{M_Z}=0.4$ with $\Lambda=1\unit{TeV}$; arrows indicate the regions accessible to CDMS II (brown [dark gray]) and SuperCDMS (green [light gray]).  The right column shows how the allowed (unshaded) regions are further constrained by increasing the cutoff scale to $\Lambda=10^9\unit{GeV}$.  For $M_h=120\unit{GeV}$, intersection with the vacuum stability bounds due to large $h$ minima removes the allowed $\of{a_2\comma M_S}$ region below $a_2\simeq 0.25$.  For $M_h=200\unit{GeV}$, intersection with the perturbativity bounds removes the allowed regions above $a_2\simeq 0.35$.  For $M_h=350\unit{GeV}$, perturbativity limits eliminate the entire region of $\of{a_2,M_S}$ space.  These effects are consistent with Fig.~\ref{fig:mhva2_b4}.}
	\label{fig:dmplots10to9gev}
\end{figure}

We also show in Fig.~\ref{fig:dmplotmh200} various features relevant for dark matter.  Given the model parameters $M_h$, $M_S$, and $a_2$, we calculate the annihilation cross section $\sigma_{ann}$ and relic density $\Omega_S h^2$ for singlet cold dark matter (CDM), following the discussions in Refs.~\cite{He:2008qm,Burgess:2000yq}\footnote{We have also made use of the Higgs decay width program HDECAY of Ref.~\cite{Djouadi:1997yw}.}.  The blue band in Fig.~\ref{fig:dmplotmh200} shows the regions required to obtain a relic density consistent with the WMAP range\footnote{These bands in Figs.~\ref{fig:dmplotmh200}, \ref{fig:dmplots1tev}, and \ref{fig:dmplots10to9gev} are reproduced from Ref.~\cite{He:2008qm} using the 90\% C.L. range for the relic density taken from the Particle Data Group \cite{Amsler:2008zzb}, $0.092\leq\Omega_{dm} h^2\leq 0.118$.}, assuming that the singlet saturates the CDM relic density.  We also calculate the cross section for direct detection of singlet dark matter by elastic scattering off a nucleon, $\sigma_{dd}$, according to Refs.~\cite{Carroll:2009dw,He:2008qm}; this cross section is then scaled by the fraction of the relic density constituted by the singlet, 
\begin{equation*}
	\tilde{\sigma}_{dd}=\sigma_{dd}\ \frac{\Omega_S h^2}{\Omega_{dm} h^2}=\sigma_{dd}\ \frac{\Omega_S h^2}{0.105} \leq \sigma_{dd}\ \ \ ,
\end{equation*}
as in Ref.~\cite{Barger:2007im}, to account for reduced flux in detector experiments in scenarios where the singlet does not saturate the CDM relic density.  The computed $\tilde{\sigma}_{dd}$ is then compared to the experimental bounds\footnote{See Ref.~\cite{web_dmtools} for a compilation of dark matter detection limits from many different sources.  We thank M.~McCaskey for providing numerical dark matter detection limits as a function of the dark matter mass.} of XENON10 \cite{Angle:2007uj} and CDMS-II \cite{Ahmed:2008eu} as well as the projected sensitivity of SuperCDMS \cite{Brink:2005ej}.  We show in Fig.~\ref{fig:dmplotmh200} the regions of $\of{a_2\comma M_S}$ parameter space excluded by CDMS-II (brown arrow) and the regions to which SuperCDMS will be sensitive (green arrow).  We do not show the exclusions from XENON10 because they are superseded by the more restrictive CDMS-II limits.

We plot the vacuum stability regions for a set of representative values of $M_h$ and $b_4\of{M_Z}$ in Fig.~\ref{fig:dmplots1tev}.  Each row of plots corresponds to a single Higgs mass (top row, $M_h=120\unit{GeV}$; middle row, $M_h=200\unit{GeV}$; bottom row, $M_h=350\unit{GeV}$) while each column corresponds to a single value of the singlet self-coupling at the input scale (left column, $b_4\of{M_Z}=0.001$; middle column, $b_4\of{M_Z}=0.4$; right column, $b_4\of{M_Z}=0.8$).  A cutoff scale of $\Lambda = 1\unit{TeV}$ is used.  The red curves of Fig.~\ref{fig:dmplots1tev} mark the boundary between the regions with stable EW/DM minima (unshaded lower areas) and regions which are excluded due to the presence of unviable $S\neq 0$ minima (shaded upper areas).  The blue bands, as in Fig.~\ref{fig:dmplotmh200}, outline the regions consistent with the observed relic density, while the brown and green arrows indicate the regions of CDMS-II exclusion and SuperCDMS sensitivity, respectively.  

\begin{figure}
	\includegraphics{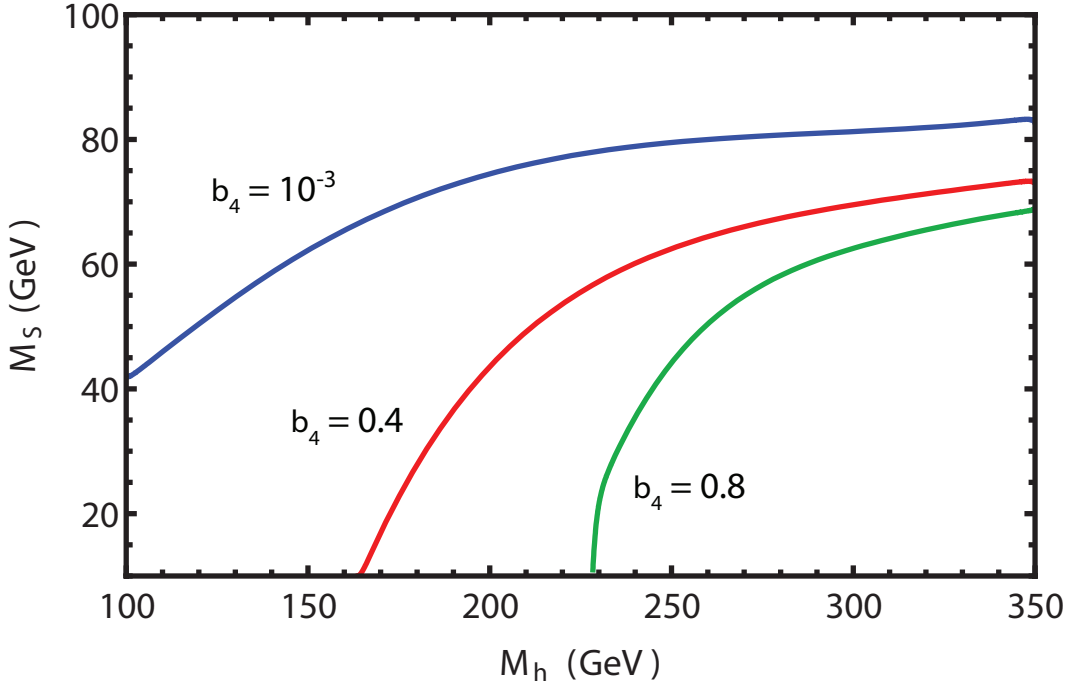}
	\caption{The minimum singlet mass compatible with both the WMAP relic density constraints (assuming the singlet saturates the relic density) and vacuum stability/perturbativity as a function of the Higgs mass, for three representative values of the singlet quartic self-coupling $b_4\of{M_Z}$.  The cutoff scale is taken to be $1\unit{TeV}$.}
	\label{fig:msmin}
\end{figure}

Fig.~\ref{fig:dmplots1tev} demonstrates that $b_4\of{M_Z}$ must be sufficiently large to have a stable electroweak/dark matter minimum for a given $M_S\comma a_2\comma\text{and}\ M_h$.  In other words, increasing $b_4\of{M_Z}$ from the left plots to the right plots of Fig.~\ref{fig:dmplots1tev} permits larger values of $a_2$ for a given $M_S$ and $M_h$.  There are two implications of this.  First, for sufficiently small $b_4$, one may encounter a minimum value of $M_S$ -- given by the intersection of a given line with the blue band -- for which the $\of{h=v_0\comma S=0}$ vacuum is stable and for which $S$ saturates the CDM relic density.  Increasing the Higgs mass for a fixed $b_4\of{M_Z}$ also increases the minimum singlet mass allowed by vacuum stability, as evidenced by moving top to bottom down a column of plots in Fig.~\ref{fig:dmplots1tev}.  We show this effect in more detail in Fig.~\ref{fig:msmin} where we plot the minimum $M_S$ as a function of the Higgs mass for the same representative values of $b_4\of{M_Z}$.  

However, we see from Fig.~\ref{fig:dmplots1tev} that except in cases of larger $M_h$ and very small $b_4\of{M_Z}$ the experimental limits on the singlet mass are more restrictive than the combined theoretical vacuum stability and relic density limits shown in Fig.~\ref{fig:msmin}.  For a relatively low cutoff scale, {\em i.e.} $\Lambda=1\unit{TeV}$, it is unlikely that discovery of both a Higgs and dark matter scalar, and measurement of their masses, would divulge information on the $S$ self-coupling by way of Fig.~\ref{fig:msmin} in which the $S$ is assumed to saturate the relic density.

\begin{figure}
	\includegraphics{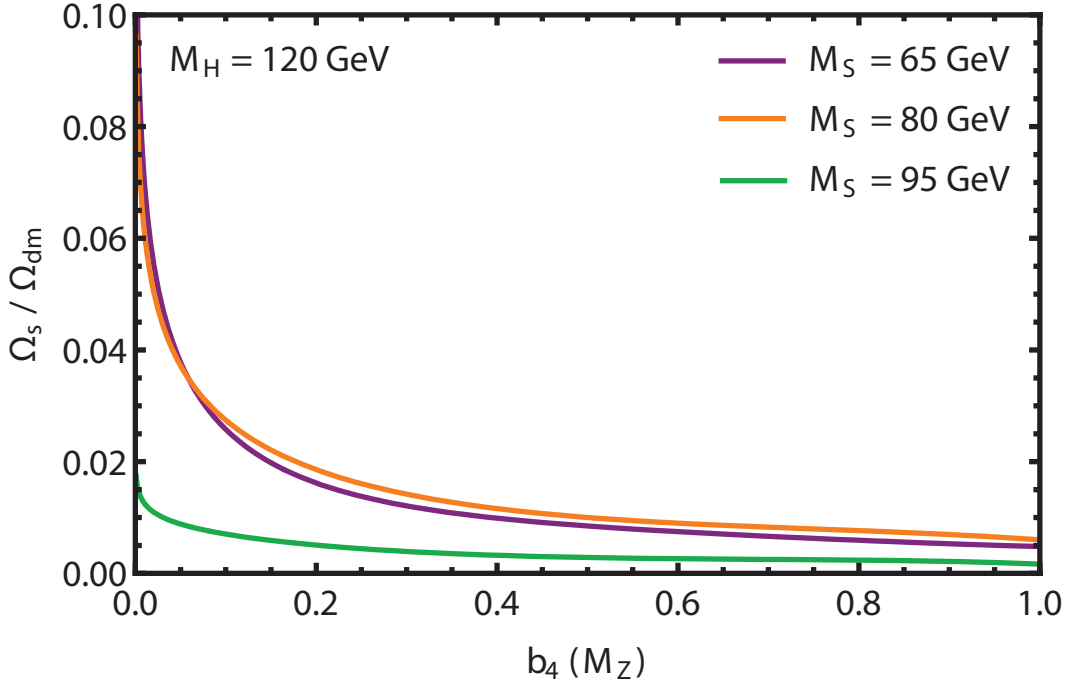}
	\caption{The minimum singlet relic density, as a fraction of the total relic density, vs. $b_4\of{M_Z}$ for $M_S=$ 65 (purple [dark gray] curve), 80 (orange [light gray] curve), and 95 $\unit{GeV}$ (green [medium gray] curve) and taking $M_h=120\unit{GeV}$ and $\Lambda=1\unit{TeV}$.  The regions below the respective curves are disallowed because they correspond to larger $a_2$ values which are excluded when requiring vacuum stability up to $\Lambda=1\unit{TeV}$.}
	\label{fig:relicdens}
\end{figure}

The second implication is the existence of vacuum stability bounds on the singlet relic density as a function of the quartic coupling $b_4$ for scenarios in which the singlet is no longer required to saturate the relic density.  Fig.~\ref{fig:dmplots1tev} shows that there is a maximum $a_2$ -- corresponding to the red curves -- allowed by vacuum stability when keeping the other parameters $\of{M_S\comma M_h\comma b_4\of{M_Z}\comma\Lambda}$ fixed.  The singlet relic density is inversely proportional to the annihilation cross section which itself is proportional to $a_2^2$ \cite{He:2008qm,Burgess:2000yq,Bento:2001yk}; thus, the maximum value of $a_2$ allowed by vacuum stability gives a minimum relic density for the singlet.  We plot in Fig.~\ref{fig:relicdens} this minimum $S$ relic density, as a fraction of the total observed relic density ($\Omega_{dm} h^2=0.105$ \cite{Amsler:2008zzb}), vs. $b_4\of{M_Z}$.  We use a cutoff scale of $1\unit{TeV}$, the Higgs mass is fixed at $120\unit{GeV}$, and we choose three fixed values for the singlet mass, $M_S=$ 65 (purple curve), 80 (orange curve), and 95~GeV (green curve), all of which fall in the region not excluded by CDMS-II but accessible by SuperCDMS (see the top row of Fig.~\ref{fig:dmplots1tev}).  We observe that a scalar singlet with a mass in the vicinity of $65-80\unit{GeV}$ and constituting 2\% of the relic density would require a quartic self-coupling $b_4\of{M_Z}\geq 0.2$ to be consistent with vacuum stability.  Measurement of $M_S$ and the singlet's coupling with the Higgs (or determination of its relic density by other means) could therefore provide enough information to constrain the singlet's self-interaction features from theoretical considerations of vacuum stability if $S$ is only a minor constituent of the total dark matter relic density.

\begin{figure}
	\includegraphics{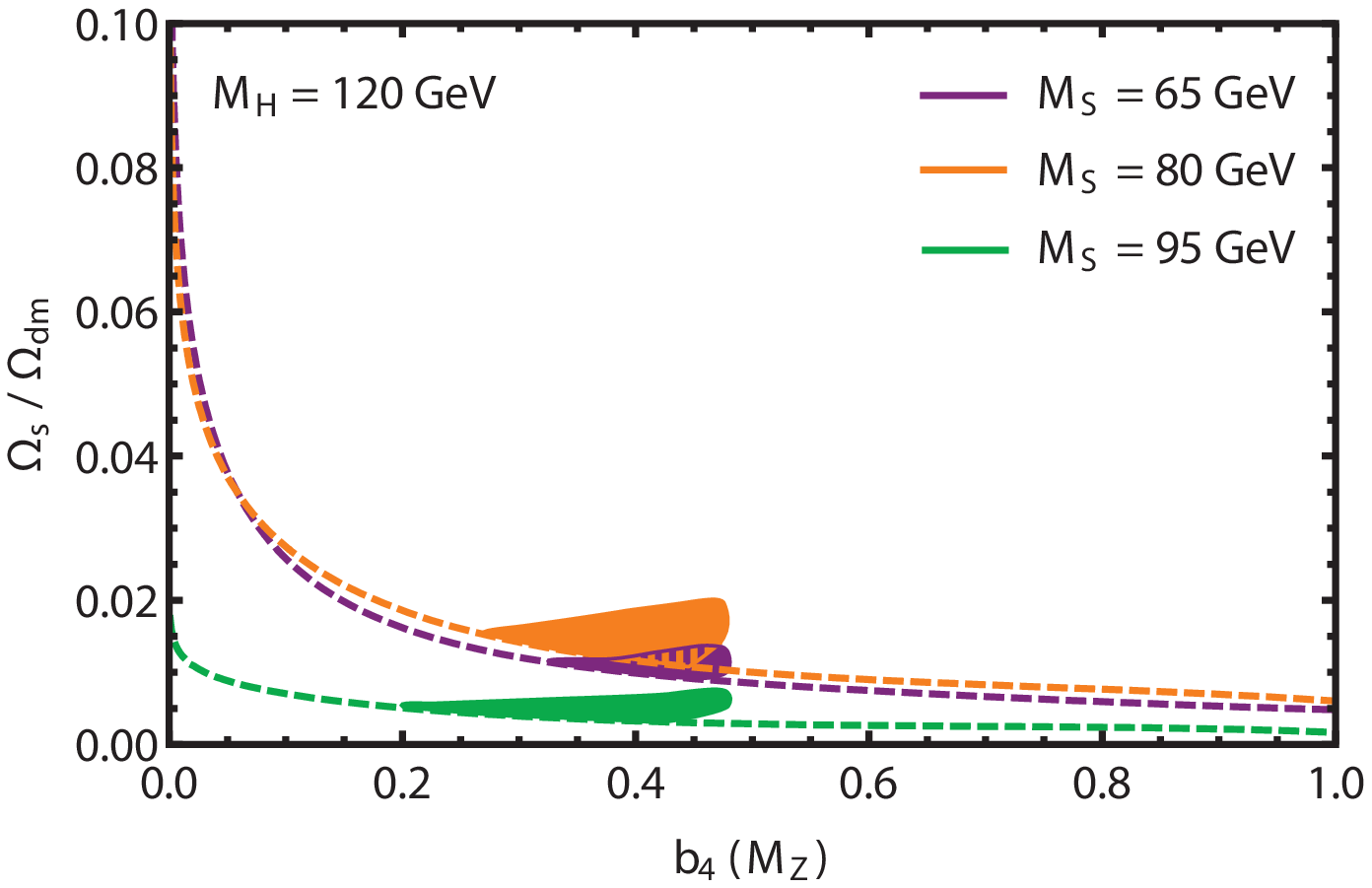}
	\caption{The dashed lines show the minimum fractional singlet relic densities from Fig.~\ref{fig:relicdens} vs. $b_4\of{M_Z}$ when requiring vacuum stability to $1\unit{TeV}$.  Increasing the cutoff scale to $\Lambda=10^9\unit{GeV}$ severely restricts the allowed regions of $a_2$, {\em i.e.,} the singlet relic density, and $b_4$: only the small colored regions are allowed due to vacuum stability and perturbativity restrictions.  $M_h=120\unit{GeV}$ and $M_S=$ 65 (purple [dark gray]), 80 (orange [light gray]), and 95 $\unit{GeV}$ (green [medium gray]) as in Fig.~\ref{fig:relicdens}.}
	\label{fig:relicdens10to9}
\end{figure}

Finally, we comment on the effect of increasing the cutoff scale.  In the left column of Fig.~\ref{fig:dmplots10to9gev} we show again the allowed regions of $\of{a_2\comma M_S}$ space with $b_4\of{M_Z}=0.4$ (the middle column of Fig.~\ref{fig:dmplots1tev}) for a $1\unit{TeV}$ cutoff.  In the right column of Fig.~\ref{fig:dmplots10to9gev}, we show the same parameter spaces but now we require stability and perturbativity of the potential up to $\Lambda=10^9\unit{GeV}$.   For $M_h=120\unit{GeV}$ (top right), the $\of{a_2\comma M_S}$ region below $a_2\simeq 0.25$ is disallowed due to the occurrence of large $h$ minima.  For $M_h=200\unit{GeV}$ (middle right), the perturbativity bounds remove the allowed regions above $a_2\simeq 0.35$, while the perturbativity bounds prohibit all of the $\of{a_2\comma M_S}$ parameter space shown for $M_h=350\unit{GeV}$ (bottom right).  These restrictions are on $a_2$ only since they are independent of the singlet mass for the cases shown; they are consistent with Fig.~\ref{fig:mhva2_b4} from which we can determine the values of $a_2$ that allow stability and perturbativity of the potential up to a $10^9\unit{GeV}$ cutoff once the Higgs mass is fixed\footnote{Recall that the less restrictive choice of perturbativity constraints corresponds to the combined solid-colored and striped regions of Fig.~\ref{fig:mhva2_b4}.}. 

More restrictive bounds on the relic density of the singlet can be obtained by requiring stability and perturbativity of the potential to higher scales.  For example, the existence of a minimum $a_2$ coupling in the top right plot of Fig.~\ref{fig:dmplots10to9gev} implies a maximum singlet relic density less than the total relic density.  We show in Fig.~\ref{fig:relicdens10to9} another plot of the fraction $\Omega_S/\Omega_{dm}$ vs. $b_4\of{M_Z}$ for $M_h=120\unit{GeV}$ and $M_S=\of{65\comma 80\comma 95}\unit{GeV}$ (purple, orange, and green, respectively), but now we require stability and perturbativity of the potential up to $\Lambda=10^9\unit{GeV}$.  Whereas in Fig.~\ref{fig:relicdens} the regions above the curves (all the way up to $\Omega_S/\Omega_{dm}=1$) were allowed for $\Lambda=1\unit{TeV}$, increasing the cutoff scale to $\Lambda=10^9\unit{GeV}$ greatly diminishes the allowed regions to the small colored areas of Fig.~\ref{fig:relicdens10to9} (the $1\unit{TeV}$ restrictions are shown as dashed curves in Fig.~\ref{fig:relicdens10to9} for comparison).  Assuming a Higgs mass of 120~GeV, we observe that imposing stability and perturbativity of the effective potential up to a $10^9\unit{GeV}$ cutoff scale restricts the singlet to scenarios in which it constitutes 2\% or less of the CDM relic density and has a moderate (0.2 to 0.5 in our normalization) self-interaction coupling.

\section{Conclusions}
\label{sec:concl}

We have performed a study of the \model~using requirements of vacuum stability and perturbativity of the one-loop effective potential.  The extra scalar degree of freedom introduced in this model can provide a suitable cold dark matter candidate.  In this scenario, the relevant parameters, in addition to the Higgs mass, are the singlet mass $M_S$, its $a_2$ coupling to the Higgs by the $H^\dagger HS^2$ term of the potential, and the quartic self-coupling $b_4$.  $M_S$ and $a_2$ govern the annihilation cross section of the singlet, its relic density, and its cross section for scattering off a nucleon.  By imposing the requirements of vacuum stability and perturbativity of the potential, we are able to constrain the parameter space of the \model.  If the singlet saturates the relic density, stability of the $\of{h=v_0\comma S=0}$ minimum of the potential up to the 1~TeV scale necessitates a lower bound on $M_S$ that varies with $b_4$.  Measurement of $M_S$ could therefore in principle constrain the singlet self-coupling; however, current experimental limits already exclude a wide range of singlet masses, thus limiting our ability to constrain the self-interactions.  If the singlet undersaturates the relic density, we can again place limits on $b_4$ from vacuum stability; however, for values of $M_S$ not already excluded by experiment, such limits are relevant when using a 1~TeV cutoff for our theory only if the singlet constitutes $\mathcal{O}\of{1\%}$ of the total relic density.  Increasing the cutoff scale places much stronger bounds on the singlet relic density and its self-coupling.  For instance, a scalar singlet discovered by SuperCDMS would compose at most 2\% of the relic density and have a moderate self-coupling in the range $0.2-0.5$ if we require stability and perturbativity of the effective potential up to $10^9\unit{GeV}$.

These analyses of the \model~parameter space just described depend on the as-yet undiscovered Higgs mass; we have also studied the reverse effect in which the \model~parameters affect bounds on the Higgs mass arising from vacuum stability and perturbativity.  Our numerical studies found that triviality, {\em i.e.,} avoidance of Landau poles below the cutoff scale, is an insufficient criterion for determining upper bounds on the Higgs mass, forcing us to require perturbativity of the running couplings instead.  Our analysis showed that increasing the $a_2$ coupling decreases both vacuum stability and perturbativity bounds on the Higgs mass, while increasing $b_4$ decreases only the upper bounds from perturbativity.  Additionally, we have shown that requiring vacuum stability and perturbativity of the potential for larger $a_2$ and $b_4$ can force the appearance of new physics at a scale below the Planck scale.

In the case of the \model, theoretical requirements of vacuum stability and perturbativity have interesting implications for the relationships between the singlet scalar self-interaction, the corresponding fraction of the dark matter relic density, and the scale at which one expects additional degrees of freedom to become relevant. These considerations can also affect the theoretical interpretation of searches for the SM Higgs boson and for scalar dark matter in direct detection experiments. It is natural to ask if there exist similar implications for other non-singlet scalar dark matter scenarios, such as those in which additional scalar degrees of freedom have non-trivial gauge quantum numbers (e.g., the triplet model discussed in Ref.~\cite{FileviezPerez:2008bj}). Also of interest are scenarios in which extra scalars can contribute to a strong first-order electroweak phase transition. Analyses for these situations are the subjects of ongoing work.

\section*{Acknowledgements}
This work was supported in part by Department of Energy contract
DE-FG02-08ER41531 and the
Wisconsin Alumni Research Foundation. We thank P.~Langacker, D.~J.~H.~Chung, and D.~Hooper for useful discussions and M.~McCaskey for providing the numerical dark matter detection limits as a function of the dark matter mass. We thank M.~Sher for a careful reading of this manuscript and useful comments.  MJRM also thanks NORDITA, the Aspen Center for Physics, and the TRIUMF Theory Group where part of this work was completed. 

\appendix 

\section{Notation \& Normalization of the Potential}
\label{appendix:notation}
Here we compare our normalization choices for the parameters of the potential, Eq.~\eqnref{Scalar_singlet_tree_level_potential}, with other choices found in the literature.  In the convention commonly found in the literature ({\em e.g.} Refs.~\cite{Hambye:1996wb, Lerner:2009xg, Clark:2009dc, Profumo:2007wc}) where $m_h^2=2\lambda v_0^2$, the coupling $\lambda$ is a factor of six smaller than our coupling $\lambda$.  Our normalization choice of $m_h^2=\lambda v_0^2/3$ follows that of Ref.~\cite{Ford:1992mv} which was our source for the SM RGEs; this choice leads to a Higgs quartic self-coupling term in the potential preceded by $\frac{1}{4!}$, thus mimicking conventional $\phi^4$ theory.  Additionally, our Higgs-singlet couplings $a_1$ and $a_2$ are a factor of two larger than those found in Refs.~\cite{Profumo:2007wc}, but smaller by a factor of two compared to Ref.~\cite{Lerner:2009xg} (in which the authors use the notation $\lambda_{hs}$ rather than $a_2$ and do not include an $a_1$ type term because of their choice of a $Z_2$ symmetry from the outset).

\section{One-loop Potential Particle Content}
\label{appendix:1Lpot}
The numerical factors and field-dependent masses in Eq.~\eqnref{One_loop_potential_formula} are as follows for the indicated particles:
\begin{align*}
	&j = W & n_W &= 6 & c_W &= \frac{5}{6} & m_W^2 &= \frac{1}{4}g^2h^2\ \ ,\\\nonumber 
	&j = Z & n_Z &= 3 & c_Z &= \frac{5}{6} & m_Z^2 &= \frac{1}{4}\of{g^2+{g'}^2}h^2\ \ ,\\\nonumber 
	&j = t & n_t &= -12 & c_t &= \frac{3}{2} & m_t^2 &= \frac{1}{2}y_t^2h^2\ \ ,\\\nonumber 
	&j = \phi & n_\phi &= 3 & c_\phi &= \frac{3}{2} & m_\phi^2 &= m^2+\frac{\lambda}{6}h^2+a_2S^2\ \ ,\\\nonumber 
	&j = \pm & n_\pm &= 1 & c_\pm &= \frac{3}{2} & m_\pm^2 &=\ \text{see Eqs.~\eqnref{eq:plus_minus_mass_evals} and \eqnref{eq:msquared}}\ \ .\\\nonumber 
\end{align*}
Here, $\phi$ stands for the would-be Goldstone bosons and $\pm$ stands for the diagonalized Higgs/singlet contributions.  We have not included contributions to the one-loop potential from fermions other than the top quark due to the smallness of their Yukawa couplings.

\section{One-loop Beta Functions}
\label{appendix:beta}

We give here the $\beta$ and $\gamma$ functions for all parameters of the \model.  All couplings depend on the 't Hooft scale $\mu$ which we have suppressed for clarity.

\begin{align}
	&\beta_{g}=\loopfac\lp -\frac{19}{6}g^3\rp\\
	&\beta_{g'}=\loopfac\lp \frac{41}{6}{g'}^3\rp\\
	&\beta_{g_3}=\loopfac\lp -7g_3^3\rp\\
	&\beta_{y_t}=\loopfac\lp \frac{9}{2}y_t^3 - 8g_3^2 y_t -\frac{9}{4}g^2 y_t - \frac{17}{12}{g'}^2 y_t\rp\\
	&\beta_\lambda =\loopfac\lp 4\lambda^2 + 12a_2^2 - 36y_t^4 + 12\lambda y_t^2 - 9\lambda g^2 -3\lambda {g'}^2 + \frac{9}{4}{g'}^4
		+ \frac{9}{2}g^2{g'}^2 + \frac{27}{4}g^4\rp\label{Lambda_beta_function}\\
	&\beta_{a_2}=\loopfac\lp 8a_2^2 + 6a_2b_4 - \frac{9}{2}a_2g^2 - \frac{3}{2}a_2{g'}^2 + 6a_2y_t^2 + 2a_2\lambda\rp\label{eq:a2_beta_function}\\
	&\beta_{b_4}=\loopfac\lp 8a_2^2 + 18b_4^2\rp\\
	&\beta_{m^2}=\loopfac\lp m^2\lb 2\lambda + 6y_t^2 - \frac{9}{2}g^2 - \frac{3}{2}{g'}^2\rb + 2a_2b_2\rp\\
	&\beta_{b_2}=\loopfac\lp 8m^2a_2 + 6b_2b_4\rp\\
	&\gamma_h =\loopfac\lp 3y_t^2 - \frac{9}{4}g^2 - \frac{3}{4}{g'}^2\rp\\
	&\gamma_S =0\\
	&\mu\frac{d\Omega}{d\mu}=\loopfac\lp 2m^4+\frac{1}{2}b_2^2\rp
\end{align}

\bibliography{VacStab06}

\providecommand{\href}[2]{#2}\begingroup\raggedright\begin{thebibliography}{10}

\bibitem{Barger:2008jx}
V.~Barger, P.~Langacker, M.~McCaskey, M.~Ramsey-Musolf, and G.~Shaughnessy,
  ``{Complex Singlet Extension of the Standard Model},''
  \href{http://dx.doi.org/10.1103/PhysRevD.79.015018}{{\em Phys. Rev.} {\bf
  D79} (2009)  015018},
\href{http://arxiv.org/abs/0811.0393}{{\tt arXiv:0811.0393 [hep-ph]}}.

\bibitem{Bento:2001yk}
M.~C. Bento, O.~Bertolami, and R.~Rosenfeld, ``{Cosmological constraints on an
  invisibly decaying Higgs boson},''
  \href{http://dx.doi.org/10.1016/S0370-2693(01)01078-4}{{\em Phys. Lett.} {\bf
  B518} (2001)  276--281},
\href{http://arxiv.org/abs/hep-ph/0103340}{{\tt arXiv:hep-ph/0103340}}.

\bibitem{Silveira:1985rk}
V.~Silveira and A.~Zee, ``{SCALAR PHANTOMS},''
\href{http://dx.doi.org/10.1016/0370-2693(85)90624-0}{{\em Phys. Lett.} {\bf
  B161} (1985)  136}.

\bibitem{McDonald:1993ex}
J.~McDonald, ``{Gauge Singlet Scalars as Cold Dark Matter},''
  \href{http://dx.doi.org/10.1103/PhysRevD.50.3637}{{\em Phys. Rev.} {\bf D50}
  (1994)  3637--3649},
\href{http://arxiv.org/abs/hep-ph/0702143}{{\tt arXiv:hep-ph/0702143}}.

\bibitem{Burgess:2000yq}
C.~P. Burgess, M.~Pospelov, and T.~ter Veldhuis, ``{The minimal model of
  nonbaryonic dark matter: A singlet scalar},''
  \href{http://dx.doi.org/10.1016/S0550-3213(01)00513-2}{{\em Nucl. Phys.} {\bf
  B619} (2001)  709--728},
\href{http://arxiv.org/abs/hep-ph/0011335}{{\tt arXiv:hep-ph/0011335}}.

\bibitem{He:2008qm}
X.-G. He, T.~Li, X.-Q. Li, J.~Tandean, and H.-C. Tsai, ``{Constraints on Scalar
  Dark Matter from Direct Experimental Searches},''
  \href{http://dx.doi.org/10.1103/PhysRevD.79.023521}{{\em Phys. Rev.} {\bf
  D79} (2009)  023521},
\href{http://arxiv.org/abs/0811.0658}{{\tt arXiv:0811.0658 [hep-ph]}}.

\bibitem{McDonald:2001vt}
J.~McDonald, ``{Thermally generated gauge singlet scalars as self- interacting
  dark matter},'' \href{http://dx.doi.org/10.1103/PhysRevLett.88.091304}{{\em
  Phys. Rev. Lett.} {\bf 88} (2002)  091304},
\href{http://arxiv.org/abs/hep-ph/0106249}{{\tt arXiv:hep-ph/0106249}}.

\bibitem{Bento:2000ah}
M.~C. Bento, O.~Bertolami, R.~Rosenfeld, and L.~Teodoro, ``{Self-interacting
  dark matter and invisibly decaying Higgs},''
  \href{http://dx.doi.org/10.1103/PhysRevD.62.041302}{{\em Phys. Rev.} {\bf
  D62} (2000)  041302},
\href{http://arxiv.org/abs/astro-ph/0003350}{{\tt arXiv:astro-ph/0003350}}.

\bibitem{Barger:2007im}
V.~Barger, P.~Langacker, M.~McCaskey, M.~J. Ramsey-Musolf, and G.~Shaughnessy,
  ``{LHC Phenomenology of an Extended Standard Model with a Real Scalar
  Singlet},'' \href{http://dx.doi.org/10.1103/PhysRevD.77.035005}{{\em Phys.
  Rev.} {\bf D77} (2008)  035005},
\href{http://arxiv.org/abs/0706.4311}{{\tt arXiv:0706.4311 [hep-ph]}}.

\bibitem{O'Connell:2006wi}
D.~O'Connell, M.~J. Ramsey-Musolf, and M.~B. Wise, ``{Minimal Extension of the
  Standard Model Scalar Sector},''
  \href{http://dx.doi.org/10.1103/PhysRevD.75.037701}{{\em Phys. Rev.} {\bf
  D75} (2007)  037701},
\href{http://arxiv.org/abs/hep-ph/0611014}{{\tt arXiv:hep-ph/0611014}}.

\bibitem{Profumo:2007wc}
S.~Profumo, M.~J. Ramsey-Musolf, and G.~Shaughnessy, ``{Singlet Higgs
  Phenomenology and the Electroweak Phase Transition},''
  \href{http://dx.doi.org/10.1088/1126-6708/2007/08/010}{{\em JHEP} {\bf 08}
  (2007)  010},
\href{http://arxiv.org/abs/0705.2425}{{\tt arXiv:0705.2425 [hep-ph]}}.

\bibitem{Espinosa:1993bs}
J.~R. Espinosa and M.~Quiros, ``{The Electroweak phase transition with a
  singlet},'' \href{http://dx.doi.org/10.1016/0370-2693(93)91111-Y}{{\em Phys.
  Lett.} {\bf B305} (1993)  98--105},
\href{http://arxiv.org/abs/hep-ph/9301285}{{\tt arXiv:hep-ph/9301285}}.

\bibitem{Anderson:1991zb}
G.~W. Anderson and L.~J. Hall, ``{The Electroweak phase transition and
  baryogenesis},''
\href{http://dx.doi.org/10.1103/PhysRevD.45.2685}{{\em Phys. Rev.} {\bf D45}
  (1992)  2685--2698}.

\bibitem{Eboli:2000ze}
O.~J.~P. Eboli and D.~Zeppenfeld, ``{Observing an invisible Higgs boson},''
  \href{http://dx.doi.org/10.1016/S0370-2693(00)01213-2}{{\em Phys. Lett.} {\bf
  B495} (2000)  147--154},
\href{http://arxiv.org/abs/hep-ph/0009158}{{\tt arXiv:hep-ph/0009158}}.

\bibitem{Barger:2006sk}
V.~Barger, P.~Langacker, and G.~Shaughnessy, ``{Collider signatures of singlet
  extended Higgs sectors},''
  \href{http://dx.doi.org/10.1103/PhysRevD.75.055013}{{\em Phys. Rev.} {\bf
  D75} (2007)  055013},
\href{http://arxiv.org/abs/hep-ph/0611239}{{\tt arXiv:hep-ph/0611239}}.

\bibitem{Davoudiasl:2004aj}
H.~Davoudiasl, T.~Han, and H.~E. Logan, ``{Discovering an invisibly decaying
  Higgs at hadron colliders},''
  \href{http://dx.doi.org/10.1103/PhysRevD.71.115007}{{\em Phys. Rev.} {\bf
  D71} (2005)  115007},
\href{http://arxiv.org/abs/hep-ph/0412269}{{\tt arXiv:hep-ph/0412269}}.

\bibitem{Brink:2005ej}
{\bf CDMS-II} Collaboration, P.~L. Brink {\em et al.}, ``{Beyond the CDMS-II
  dark matter search: SuperCDMS},''
\href{http://arxiv.org/abs/astro-ph/0503583}{{\tt arXiv:astro-ph/0503583}}.

\bibitem{Carroll:2009dw}
S.~M. Carroll, S.~Mantry, and M.~J. Ramsey-Musolf, ``{Implications of a Scalar
  Dark Force for Terrestrial Experiments},''
\href{http://arxiv.org/abs/0902.4461}{{\tt arXiv:0902.4461 [hep-ph]}}.

\bibitem{McDonald:2007ka}
J.~McDonald, N.~Sahu, and U.~Sarkar, ``{Seesaw at Collider, Lepton Asymmetry
  and Singlet Scalar Dark Matter},''
  \href{http://dx.doi.org/10.1088/1475-7516/2008/04/037}{{\em JCAP} {\bf 0804}
  (2008)  037},
\href{http://arxiv.org/abs/0711.4820}{{\tt arXiv:0711.4820 [hep-ph]}}.

\bibitem{Spergel:1999mh}
D.~N. Spergel and P.~J. Steinhardt, ``{Observational evidence for
  self-interacting cold dark matter},''
  \href{http://dx.doi.org/10.1103/PhysRevLett.84.3760}{{\em Phys. Rev. Lett.}
  {\bf 84} (2000)  3760--3763},
\href{http://arxiv.org/abs/astro-ph/9909386}{{\tt arXiv:astro-ph/9909386}}.

\bibitem{Hambye:1996wb}
T.~Hambye and K.~Riesselmann, ``{Matching conditions and Higgs mass upper
  bounds revisited},'' \href{http://dx.doi.org/10.1103/PhysRevD.55.7255}{{\em
  Phys. Rev.} {\bf D55} (1997)  7255--7262},
\href{http://arxiv.org/abs/hep-ph/9610272}{{\tt arXiv:hep-ph/9610272}}.

\bibitem{Casas:1996aq}
J.~A. Casas, J.~R. Espinosa, and M.~Quiros, ``{Standard Model stability bounds
  for new physics within LHC reach},''
  \href{http://dx.doi.org/10.1016/0370-2693(96)00682-X}{{\em Phys. Lett.} {\bf
  B382} (1996)  374--382},
\href{http://arxiv.org/abs/hep-ph/9603227}{{\tt arXiv:hep-ph/9603227}}.

\bibitem{Clark:2009dc}
T.~E. Clark, B.~Liu, S.~T. Love, and T.~ter Veldhuis, ``{The Standard Model
  Higgs Boson-Inflaton and Dark Matter},''
\href{http://arxiv.org/abs/0906.5595}{{\tt arXiv:0906.5595 [hep-ph]}}.

\bibitem{Lerner:2009xg}
R.~N. Lerner and J.~McDonald, ``{Gauge singlet scalar as inflaton and thermal
  relic dark matter},''
\href{http://arxiv.org/abs/0909.0520}{{\tt arXiv:0909.0520 [hep-ph]}}.

\bibitem{Ford:1992mv}
C.~Ford, D.~R.~T. Jones, P.~W. Stephenson, and M.~B. Einhorn, ``{The Effective
  potential and the renormalization group},''
  \href{http://dx.doi.org/10.1016/0550-3213(93)90206-5}{{\em Nucl. Phys.} {\bf
  B395} (1993)  17--34},
\href{http://arxiv.org/abs/hep-lat/9210033}{{\tt arXiv:hep-lat/9210033}}.

\bibitem{Sher:1988mj}
M.~Sher, ``{Electroweak Higgs Potentials and Vacuum Stability},''
\href{http://dx.doi.org/10.1016/0370-1573(89)90061-6}{{\em Phys. Rept.} {\bf
  179} (1989)  273--418}.

\bibitem{Lindner:1985uk}
M.~Lindner, ``{Implications of Triviality for the Standard Model},''
\href{http://dx.doi.org/10.1007/BF01479540}{{\em Zeit. Phys.} {\bf C31} (1986)
  295}.

\bibitem{Lindner:1988ww}
M.~Lindner, M.~Sher, and H.~W. Zaglauer, ``{Probing Vacuum Stability Bounds at
  the Fermilab Collider},''
\href{http://dx.doi.org/10.1016/0370-2693(89)90540-6}{{\em Phys. Lett.} {\bf
  B228} (1989)  139}.

\bibitem{Sher:1993mf}
M.~Sher, ``{Precise vacuum stability bound in the standard model},''
  \href{http://dx.doi.org/10.1016/0370-2693(93)91586-C}{{\em Phys. Lett.} {\bf
  B317} (1993)  159--163},
\href{http://arxiv.org/abs/hep-ph/9307342}{{\tt arXiv:hep-ph/9307342}}.

\bibitem{Casas:1994us}
J.~A. Casas, J.~R. Espinosa, M.~Quiros, and A.~Riotto, ``{The Lightest Higgs
  boson mass in the minimal supersymmetric standard model},''
  \href{http://dx.doi.org/10.1016/0550-3213(94)00508-C}{{\em Nucl. Phys.} {\bf
  B436} (1995)  3--29},
\href{http://arxiv.org/abs/hep-ph/9407389}{{\tt arXiv:hep-ph/9407389}}.

\bibitem{Casas:1994qy}
J.~A. Casas, J.~R. Espinosa, and M.~Quiros, ``{Improved Higgs mass stability
  bound in the standard model and implications for supersymmetry},''
  \href{http://dx.doi.org/10.1016/0370-2693(94)01404-Z}{{\em Phys. Lett.} {\bf
  B342} (1995)  171--179},
\href{http://arxiv.org/abs/hep-ph/9409458}{{\tt arXiv:hep-ph/9409458}}.

\bibitem{Ellis:2009tp}
J.~Ellis, J.~R. Espinosa, G.~F. Giudice, A.~Hoecker, and A.~Riotto, ``{The
  Probable Fate of the Standard Model},''
  \href{http://dx.doi.org/10.1016/j.physletb.2009.07.054}{{\em Phys. Lett.}
  {\bf B679} (2009)  369--375},
\href{http://arxiv.org/abs/0906.0954}{{\tt arXiv:0906.0954 [hep-ph]}}.

\bibitem{Isidori:2001bm}
G.~Isidori, G.~Ridolfi, and A.~Strumia, ``{On the metastability of the standard
  model vacuum},'' \href{http://dx.doi.org/10.1016/S0550-3213(01)00302-9}{{\em
  Nucl. Phys.} {\bf B609} (2001)  387--409},
\href{http://arxiv.org/abs/hep-ph/0104016}{{\tt arXiv:hep-ph/0104016}}.

\bibitem{Riesselmann:1996is}
K.~Riesselmann and S.~Willenbrock, ``{Ruling out a strongly-interacting
  standard Higgs model},''
  \href{http://dx.doi.org/10.1103/PhysRevD.55.311}{{\em Phys. Rev.} {\bf D55}
  (1997)  311--321},
\href{http://arxiv.org/abs/hep-ph/9608280}{{\tt arXiv:hep-ph/9608280}}.

\bibitem{Djouadi:1997yw}
A.~Djouadi, J.~Kalinowski, and M.~Spira, ``{HDECAY: A program for Higgs boson
  decays in the standard model and its supersymmetric extension},''
  \href{http://dx.doi.org/10.1016/S0010-4655(97)00123-9}{{\em Comput. Phys.
  Commun.} {\bf 108} (1998)  56--74},
\href{http://arxiv.org/abs/hep-ph/9704448}{{\tt arXiv:hep-ph/9704448}}.

\bibitem{Amsler:2008zzb}
{\bf Particle Data Group} Collaboration, C.~Amsler {\em et al.}, ``{Review of
  particle physics},''
\href{http://dx.doi.org/10.1016/j.physletb.2008.07.018}{{\em Phys. Lett.} {\bf
  B667} (2008)  1}.

\bibitem{web_dmtools}
R.~Gaitskell, V.~Mandic, and J.~Filippini.
  \url{http://dmtools.berkeley.edu/limitplots}.

\bibitem{Angle:2007uj}
{\bf XENON} Collaboration, J.~Angle {\em et al.}, ``{First Results from the
  XENON10 Dark Matter Experiment at the Gran Sasso National Laboratory},''
  \href{http://dx.doi.org/10.1103/PhysRevLett.100.021303}{{\em Phys. Rev.
  Lett.} {\bf 100} (2008)  021303},
\href{http://arxiv.org/abs/0706.0039}{{\tt arXiv:0706.0039 [astro-ph]}}.

\bibitem{Ahmed:2008eu}
{\bf CDMS} Collaboration, Z.~Ahmed {\em et al.}, ``{Search for Weakly
  Interacting Massive Particles with the First Five-Tower Data from the
  Cryogenic Dark Matter Search at the Soudan Underground Laboratory},''
  \href{http://dx.doi.org/10.1103/PhysRevLett.102.011301}{{\em Phys. Rev.
  Lett.} {\bf 102} (2009)  011301},
\href{http://arxiv.org/abs/0802.3530}{{\tt arXiv:0802.3530 [astro-ph]}}.

\bibitem{FileviezPerez:2008bj}
P.~Fileviez~Perez, H.~H. Patel, M.~J. Ramsey-Musolf, and K.~Wang, ``{Triplet
  Scalars and Dark Matter at the LHC},''
  \href{http://dx.doi.org/10.1103/PhysRevD.79.055024}{{\em Phys. Rev.} {\bf
  D79} (2009)  055024},
\href{http://arxiv.org/abs/0811.3957}{{\tt arXiv:0811.3957 [hep-ph]}}.

\end{thebibliography}\endgroup
\bibliographystyle{utcaps}

\end{document}